\documentclass[letterpaper,twocolumn,10pt]{article}
\usepackage{zhanggroup}

\usepackage{tikz}
\usepackage{amsfonts}
\usepackage{xspace} 
\usepackage{amsmath}
\usepackage{mathrsfs}
\usepackage{amssymb}
\usepackage{amsmath}
\usepackage{amsthm}
\usepackage{subcaption}
\usepackage{booktabs}
\usepackage{multirow}
\usepackage{graphicx} 
\usepackage{tcolorbox}
\usepackage{caption,subcaption}
\usepackage[linesnumbered,ruled]{algorithm2e}
\usepackage[sort&compress,numbers]{natbib}
\newcounter{example}[section]
\usepackage[absolute]{textpos}

\usepackage{hyperref} 
\hypersetup{
  colorlinks,
  linkcolor={blue!70!green},
  citecolor={green!70!blue},
  urlcolor={orange!70!red}
}

\newtheorem{definition}{Requirement}

\date{}
\usepackage{balance}

\newcommand{\firstAttack}{Chameleon\xspace}
\newcommand{\secondAttack}{Adverse Chameleon\xspace}
\newcommand{\camM}{Camouflager\xspace}
\newcommand{\mypara}[1]{\smallskip\noindent\textbf{#1:}}
\newcommand{\loss}{\varphi}
\newcommand{\lossM}{\mathcal{L}}
\newcommand{\encoder}{\mathcal{E}}
\newcommand{\decoder}{\mathcal{E}^{-1}}
\newcommand{\vLabel}{{{L}}}
\newcommand{\dLabel}{\ell}
\newcommand{\labelV}{\mathcal{Y}}
\newcommand{\labelD}{{y}}
\newcommand{\data}{x}
\newcommand{\dataset}{\mathcal{D}}
\newcommand{\feat}{\mathcal{F}}
\newcommand{\hijack}{{h}}
\newcommand{\orig}{{o}}
\newcommand{\camD}{{c}}
\newcommand{\autoencoder}{\mathcal{AE_C}}
\newcommand{\model}{\mathcal{M}}
\begin{document}
%-------------------------------------------------------------------------------

\begin{textblock}{12}(2,1)
\centering
To Appear in the 29th Network and Distributed System Security Symposium, 27 February – 3 March, 2022.
\end{textblock}

%-------------------------------------------------------------------------------
\title{Get a Model! Model Hijacking Attack Against Machine Learning Models}
%-------------------------------------------------------------------------------

%-------------------------------------------------------------------------------
\author{
{\rm Ahmed Salem}\ \ \
{\rm Michael Backes}\ \ \
{\rm Yang Zhang}\ \ \
\\
\textit{CISPA Helmholtz Center for Information Security}
}
%-------------------------------------------------------------------------------

\maketitle
%-------------------------------------------------------------------------------
\begin{abstract}
Machine learning (ML) has established itself as a cornerstone for various critical applications ranging from autonomous driving to authentication systems. However, with this increasing adoption rate of machine learning models, multiple attacks have emerged. One class of such attacks is training time attack, whereby an adversary executes their attack before or during the machine learning model training. In this work, we propose a new training time attack against computer vision based machine learning models, namely model hijacking attack. The adversary aims to hijack a target model to execute a different task than its original one without the model owner noticing. Model hijacking can cause accountability and security risks since a hijacked model owner can be framed for having their model offering illegal or unethical services. Model hijacking attacks are launched in the same way as existing data poisoning attacks. However, one requirement of the model hijacking attack is to be stealthy, i.e., the data samples used to hijack the target model should look similar to the model's original training dataset. To this end, we propose two different model hijacking attacks, namely Chameleon and Adverse Chameleon, based on a novel encoder-decoder style ML model, namely the Camouflager. Our evaluation shows that both of our model hijacking attacks achieve a high attack success rate, with a negligible drop in model utility.\footnote{Code for our experiments is available at \url{https://github.com/AhmedSalem2/Model-Hijacking}.}
\end{abstract}
%-------------------------------------------------------------------------------

%-------------------------------------------------------------------------------
\section{Introduction}
%-------------------------------------------------------------------------------

Machine learning (ML) has established itself as a cornerstone for various critical applications, such as autonomous driving, financial/banking application, and authentication systems.
Two of the most significant demands fueled by this increasing rate of machine learning adoption are the need for high computational power for training more complex machine learning models, and the need for high-quality training dataset.
Such high demands for data and computational power hinder individuals from training ML models on their own.
Instead, new training paradigms which involve multiple parties jointly building machine learning models have been proposed.
One such training paradigm is federated learning~\cite{BEGHIIKKMMOPRR19}.

However, this inclusion of new parties in the training process of ML models raises new security and privacy risks.
In other words, it creates an attack surface where an adversary can manipulate the training of an ML model.
This type of attacks is called training time attack.
Some examples in this domain include backdoor attacks~\cite{GDG17} and data poisoning attacks~\cite{BNL12}.

%-------------------------------------------------------------------------------
\subsection{Our Contributions}
%-------------------------------------------------------------------------------

\mypara{Motivation}
In this work, we propose a new training time attack against computer vision based machine learning models, namely the {\em model hijacking attack}.
Concretely, the adversary performs data poisoning to repurpose a target ML model designed for a certain task (\textit{original task}) to be able to perform a \textit{hijacking task} defined by the adversary.
This repurposing of the target model has to be done stealthily such that the target model owner does not detect it.
The model hijacking attack is a training time attack, hence the adversary needs to apply stealthiness with respect to two dimensions:
the first is to not jeopardize the target model's utility with respect to its original task; the second is to camouflage the poisoning data to look similar to data from the same distribution of the target model's training dataset.

Using the model hijacking attack, the adversary can hijack a target model to perform an unintended ML task, without the model's owner noticing.
This can cause accountability risks for the model owner, since now the model owner can be framed of having their own model offering illegal or unethical services.
For example, an adversary can hijack a benign image classifier into a facial recognition model for pornography movies, or even classifying such movies into different categories.
A different fairness violating scenario is to use the hijacked model to classify people's sexuality using for example their facial attributes.
In short, using this attack, the adversary can hijack a model designed to be publicly available. This will result in a public model offering an illegal or unethical service under the unintended responsibility of the hijacked model's owner.

Another risk that can be caused by the model hijacking attack is parasitic computing.
An adversary can hijack a model with public free access to implement their application, instead of hosting their own model.
This can save the adversary the cost of training their own model.
However, more importantly, it saves the adversary the cost of maintaining their own ML model.
For example, deploying and hosting a model -- in Europe -- by google can cost from 0.11\$ up to 2.44\$ per hour.\footnote{\url{https://cloud.google.com/vertex-ai/pricing\#europe}}

\mypara{Methodology}
To perform the model hijacking attack, the adversary only needs the ability to poison the target model's training dataset.
This means model hijacking is applicable for any setting that is vulnerable to data poisoning, such as federated learning.
An adversary can implement the model hijacking attack by simply poisoning the target model's training dataset (\textit{original dataset}) with their own hijacking task's training dataset (\textit{hijacking dataset}). 
However, such an attempt can be easily detected as both the original and hijacking datasets can be significantly different.
Hence, we define the following requirements for a successful model hijacking attack.
First, a hijacked model should achieve good performance when predicting any sample from the original dataset (original sample) with respect to the original task, and any sample from the hijacking dataset (hijacking sample) with respect to the hijacking task.
Second, the execution of the attack should be stealthy, i.e., samples in the hijacking dataset should be camouflaged before being used to poison (query) the target (hijacked) model.

To fulfill these requirements, we propose two model hijacking attacks, namely the {\firstAttack} attack and the {\secondAttack} attack. 
To implement both attacks, we first propose the {\camM}, an encoder-decoder based model that camouflages samples in a hijacking dataset to be more stealthy.
Specifically, the {\camM} consists of two encoders. 
The first one encodes samples from the hijacking dataset.
The second one encodes samples from a dataset that the adversary wants the hijacking dataset to be visually similar,  we refer to this dataset as the \textit{hijackee dataset}.
Ideally, the hijackee dataset should come from the same distribution as the target dataset.
The outputs of the two encoders are then fed to a decoder which outputs the camouflaged samples.
These camouflaged samples should be visually similar to the hijackee samples, but semantically similar to the hijacking samples.
It is important to note that since the {\camM} is independent of the target model, it can be used when hijacking multiple target models performing a similar task.
In other words, the {\camM} is linked with the original task, not the target model.
In this way, the adversary can achieve effective parasitic computing.
We now briefly introduce the {\firstAttack} and {\secondAttack} attacks.

\mypara{The {\firstAttack} Attack}
Our first model hijacking attack, namely the {\firstAttack} attack utilizes two different losses to train the {\camM}.
The first one is the Visual Loss, which is responsible for making the {\camM}'s output, i.e., the camouflaged samples, visually similar to the hijackee samples.
The second one is the Semantic Loss which drives the camouflaged samples to be semantically similar to the hijacking samples in order to perform the hijacking task.
In addition to training the {\camM}, the {\firstAttack} attack also needs to establish a mapping between labels of the hijacking task and the original task.
To hijack a target model, the {\firstAttack} attack poisons the original dataset using the camouflaged dataset.
Finally, to execute the attack, the adversary camouflages a hijacking sample using the {\camM}, queries the camouflaged sample to the hijacked model, and maps the predicted label back to the corresponding one of the hijacking task.

\mypara{The {\secondAttack} Attack}
The {\firstAttack} attack has a strong performance when the distributions of both the hijacking and hijackee datasets are significantly different. 
However, when these two datasets are relatively similar, the {\camM} cannot achieve its expected properties.
To overcome this, we propose an advanced version of the {\firstAttack} attack, namely the {\secondAttack} attack.
The {\secondAttack} attack adds an additional loss, namely the adverse Semantic Loss, to the Visual and Semantic Losses used for the {\firstAttack} attack.
This new loss explicitly adds the constraint to distance the semantics/features of the output of the {\camM} from the hijackee samples, to alleviate the training of the hijacking task.

\mypara{Evaluation}
To demonstrate the efficacy of model hijacking, we perform experiments in different settings using three benchmark computer vision datasets including MNIST,\footnote{\url{http://yann.lecun.com/exdb/mnist/}} CIFAR-10,\footnote{\url{https://www.cs.toronto.edu/~kriz/cifar.html}} and CelebA~\cite{LLWT15}.
Our results show that the {\firstAttack} attack achieves almost a perfect performance when attacking both CIFAR-10 and CelebA based models using MNIST as the hijacking dataset.
Specifically, it achieves above $99\%$ accuracy for MNIST classification (the hijacking task) with no performance drop for CelebA classification and less than $1\%$ drop for CIFAR-10 classification (the original tasks).
For the more complex case of using CIFAR-10 and CelebA as the hijacking datasets, our {\secondAttack} achieves $58.6\%$ and $73.7\%$ accuracy with a negligible drop in performance for their original tasks, respectively. 

Abstractly, our contributions can be summarized as:
\begin{enumerate}
\setlength\itemsep{1em}
\item We propose the first model hijacking attack against machine learning models.
\item We propose the {\camM} model which camouflages the hijacking samples for stealthy model hijacking attacks.
\item Our two proposed model hijacking attacks, i.e., the {\firstAttack} and {\secondAttack} attacks, achieve strong performance in different settings.
\end{enumerate}

%-------------------------------------------------------------------------------
\subsection{Organization}
%-------------------------------------------------------------------------------

The rest of the paper is organized as follows.
\autoref{sec:pre} presents some background knowledge and our threat model.
Next, we introduce the model hijacking attack and our two attacks, i.e., the {\firstAttack} and {\secondAttack} attacks, in \autoref{sec:meth}.
We then evaluate the performance of our two different attacks in \autoref{sec:eval} and discuss the limitations of them in \autoref{sec:discuess}.
Finally, we present the related works, discuss the limitations, and conclude the paper in \autoref{sec:related}, \autoref{sec:discuess}, and \autoref{sec:conc}, respectively.

%-------------------------------------------------------------------------------
\section{Preliminaries}
\label{sec:pre}
%-------------------------------------------------------------------------------

In this section, we start by introducing machine learning classification.
Then, we briefly present the data poisoning attack, and finally, we introduce the problem statement and threat model for the model hijacking attack.

%-------------------------------------------------------------------------------
\subsection{Machine Learning Classification Setting}
\label{sec:classifcation}
%-------------------------------------------------------------------------------

A machine learning classifier aims to classify a data sample to a certain label/class.
More concretely, on the input of a data sample $\data$, the classifier/model $\model$ predicts a vector of probabilities $\labelV$.
The size of $\labelV$, i.e., $|\labelV|$, equals to the number of unique labels, and each entry $\labelD_i$ in $\labelV$ represents the confidence of the model $\model$ assigning the sample $\data$ to the label $\dLabel_i\in\vLabel$.
For simplicity, we only consider the final predicted label as the output of the model, which is the label with the maximum probability, i.e., $\model(x)=\text{argmax}_{\dLabel_i} \labelV$.
To train the model $\model$, we need to define a loss function, such as cross-entropy loss, and utilize an optimizer to minimize the empirical loss calculated over a training dataset $\dataset$.

%-------------------------------------------------------------------------------
\subsection{Data Poisoning Attack}
%-------------------------------------------------------------------------------

Data poisoning attack~\cite{BNL12,STLLXCS18,JOBLNL18,SHNSSDG18} is a training time attack against ML models.
In this setting, the adversary first needs to create a malicious dataset $\dataset_m$.
One way of creating such dataset ($\dataset_m$) is to mislabel a set of samples to wrong classes.
Next, the adversary inserts this malicious dataset to a benign training dataset $\dataset$ to create a poisoned dataset $\dataset_p$ ($\dataset_p = \dataset_m || \dataset$).
This poisoned dataset is then used to train the target model as mentioned in~\autoref{sec:classifcation}.
The goal of data poisoning is to jeopardize the accuracy (or utility) of the target model.

%-------------------------------------------------------------------------------
\subsection{Problem Statement}
\label{sec:problemStatment}
%-------------------------------------------------------------------------------

Model hijacking attack is a training time attack where the adversary poisons a target model's training dataset, such that they can hijack the model for a different task defined by themselves.
We refer to the dataset related to the target model's original task as the \textit{original dataset}, while the dataset related to the hijacking task as the \textit{hijacking dataset}.
Intuitively, we consider the model to be hijacked when an adversary can use it -- after being trained -- to perform their own hijacking task.
This hijacking task should be different from the original one of the hijacked model.
Moreover, hijacking a model should not jeopardize its performance on the original task and should be inconspicuous to the hijacked model's owner. 
We later (\autoref{section:attackPipline}) formally define the requirements of the model hijacking attack.

A successful model hijacking attack can save the adversary the cost of maintaining their own model.
Moreover, it can lead to an accountability risk, since the hijacked model's owner can be accountable for the hijacking task which can be illegal or unethical.

%-------------------------------------------------------------------------------
\subsection{Threat Model}
\label{sec:threatModel}
%-------------------------------------------------------------------------------

The model hijacking attack does not need any assumption related to the target model.
The only assumption needed is the ability to poison the target model's training dataset, which is similar to data poisoning attacks~\cite{STLLXCS18,JOBLNL18,SHNSSDG18}.
Moreover, we assume the adversary has a hijackee dataset which they rely on to create the camouflaged dataset from the hijacking dataset.
In \autoref{sec:advAttackMeth}, we will describe how to generate this camouflaged dataset.
Ideally, the hijackee dataset should have a similar visual appearance as the target model's dataset.
However, our model hijacking attack is independent of which distribution the hijackee dataset is sampled from.
It is the adversary's decision on which dataset to use to camouflage the hijacking dataset.

The model hijacking attack can be broadly applied to any real-world scenario where a model owner collects data from different parties to train their model.
One concrete example is federated learning.
More generally, the model hijacking attack can be performed in any setting that is vulnerable to data poisoning~\cite{TTGL20,BNL12,JOBLNL18,BG192,STLLXCS18,ZLDG20,SSTS20}.

%-------------------------------------------------------------------------------
\section{Model Hijacking Attack}
\label{sec:meth}
%-------------------------------------------------------------------------------

In this section, we present our different techniques for the model hijacking attack.
We start by presenting the general pipeline of the model hijacking attack, then we
clarify the difference between the model hijacking attack and other training time attacks, i.e., data poisoning and backdoor attacks.
Finally, we present two concrete realization of the model hijacking attack, namely the {\firstAttack} and {\secondAttack} attacks.

%-------------------------------------------------------------------------------
\subsection{General Attack Pipeline}
\label{section:attackPipline}
%-------------------------------------------------------------------------------

To perform the model hijacking attack, the adversary first creates a hijacking dataset.
Then, for each label in the hijacking dataset, they define a mapping to associate it to a label of the original dataset.

This mapped label will be used as the ground truth when poisoning the target model as will be shown later.

Next, the adversary poisons the target model's training dataset with the hijacking dataset and waits for the target model to be trained, i.e., hijacked.
Once the model is hijacked, to launch the attack, the adversary creates a hijacking sample and queries it to the hijacked model. 
Finally, the adversary maps (using the inverse of the same mapping performed at the initialization of the attack) the predicted output back to the corresponding label of the hijacking task.

A straightforward approach to perform the model hijacking attack is to directly poison the training dataset of the target model with the hijacking dataset.
However, the main disadvantage of this approach is that it is easily detectable since samples in the original and the hijacking datasets can be significantly different.

To overcome this limitation, we propose a more advanced model hijacking attack, where the samples used to poison the target model are visually similar to those in the original dataset.
To this end, we propose {\camM}, which is an encoder-decoder based model that \textit{camouflages} the hijacking dataset, i.e., transforms samples in the hijacking dataset to be visually similar to those in the original dataset, while maintaining each sample's original semantics.
We refer to the hijacking dataset after being camouflaged as the \emph{camouflaged dataset}.

{\camM} is trained using both a hijacking dataset and a \textit{hijackee} dataset.
As mentioned in \autoref{sec:threatModel}, samples in the hijackee dataset are visually similar to samples in the original dataset.
\camM is based on two types of losses, namely \textit{Visual Loss} and \textit{Semantic Loss}.
Visual Loss makes the {\camM}'s output, i.e., the camouflaged dataset, visually similar to the hijackee dataset.
Semantic Loss makes the camouflaged dataset semantically similar to the hijacking dataset, to be able to implement the hijacking task.

After training the {\camM}, the adversary can use it to camouflage the hijacking dataset and use the output camouflaged dataset to poison the target model.
Finally, to launch the attack, the adversary needs to first camouflage the desired hijacking sample, then query the camouflaged sample to the hijacked model. 
In the end, the adversary maps the predicted label to the one related to the hijacking task.

A successful model hijacking attack should predict any sample from the original dataset or the camouflaged dataset correctly, but any sample from the hijacking dataset, i.e., without first getting camouflaged by the {\camM}, randomly (significantly lower than the performance of the camouflaged samples).
More formally, we define the following requirements for a successful model hijacking attack:
\begin{definition}
\label{req:1}
The hijacked model should have a similar or better performance as the target model on its original task.
\end{definition}
\begin{definition}
\label{req:2}
The hijacking dataset should be camouflaged -- to the hijackee dataset -- to make the attack more stealthy.
\end{definition}
\begin{definition}
\label{req:3}
The hijacked model should correctly classify the camouflaged samples with respect to the hijacking task.
\end{definition}
\begin{definition}
\label{req:4}
To further increase the stealthiness of the model hijacking attack, the hijacked model should classify any non-camouflaged sample from the hijacking dataset randomly, i.e., significantly lower than the performance of the camouflaged samples 
\end{definition}

For clarity, we summarise the different used datasets in \autoref{table:datasetDesc}.

%-------------------------------------------------------------------------------
\subsection{Model Hijacking v.s.\ Backdooring v.s.\ Data Poisoning}
\label{sec:clarifyDiff}
%-------------------------------------------------------------------------------

We now compare our model hijacking attack with two related training time attacks, namely the data poisoning and backdoor attacks.
The model hijacking attack follows the same attacker assumption of the poisoning and backdoor attacks.
However, using the model hijacking attack, the adversary has a different objective.

On the one hand, in the data poisoning attacks~\cite{JOBLNL18,SMKID18}, the adversary tries to jeopardize the models' utility, i.e., increasing the misclassification rate, by manipulating the training of the target model.
Similarly, the backdoor attacks~\cite{GDG17,LMALZWZ18} can also have the same aim.
Moreover, in the backdoor attack, the adversary can link a trigger with specific model output.
For example, when the trigger is inserted in any input, the target model predicts a predefined -- by the adversary -- label.
This trigger can, for example, be a white square at the corner of the input for image classification models.

On the other hand, hijacking a model is to implement different -- unethical -- tasks irrespective of the original one, without being noticed by the model owner. 
For instance, the adversary can hijack a model to implement a facial recognition classifier for pornography movies or a sexuality classifier. 
Hijacking a model saves the adversary the cost of maintaining their own model. 
In other words, hijacking a model repurposes it to perform the adversary's task.
The backdoor attack can be considered a specific instance of the model hijacking one, where the adversary's task is to predict the triggered input to a specific label.
However, the adversary is free to determine the hijacking task in the hijacking attack with the only restriction of having similar or fewer labels compared to the target model's original task.

%-------------------------------------------------------------------------------
\subsection{Building Blocks}
\label{sec:buildingBlocks}
%-------------------------------------------------------------------------------

We now introduce the building blocks for our model hijacking attack.
We start with the {\camM}, then the different losses.

\begin{figure*}[!t]
\centering
\includegraphics[width=1.4\columnwidth]{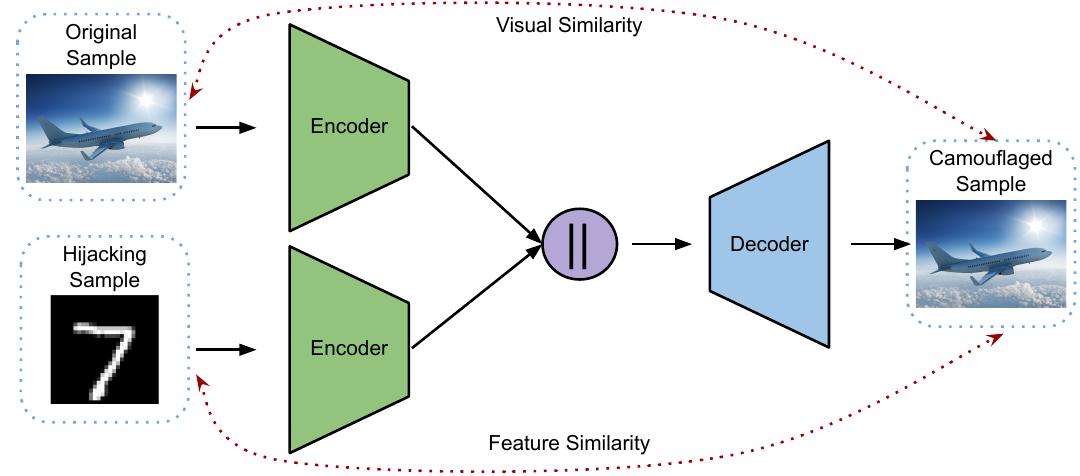}
\caption{A schematic view of the {\camM}. 
The encoders first encode the original and the hijacking samples. 
Then the outputs of the encoders are concatenated and inputted to the decoder. 
The decoder then generates the camouflaged sample which has the visual appearance of the original sample but the features of the hijacking one.}
\label{fig:camMStuct}
\end{figure*}

\mypara{{\camM} ($\autoencoder$)}
The {\camM} is an encoder-decoder based model which camouflages the hijacking dataset $\dataset_\hijack$ into the hijackee dataset $\dataset_\orig$.
We visualize the structure of the {\camM} in \autoref{fig:camMStuct}.
As the figure shows, the {\camM} consists of two encoders and one decoder.
The first encoder ($\encoder_\orig$) takes a sample ($\data_\orig$) from the hijackee dataset as its input, while the other encoder ($\encoder_\hijack$) takes a sample ($\data_\hijack$) from the hijacking dataset.
The outputs of the two encoders are then concatenated to create the input for the decoder ($\decoder$).
The decoder then generates a camouflaged sample $\data_\camD$ which has visual appearance of $\data_\orig$ with the features/semantics of $\data_\hijack$.
More formally,
\[
\autoencoder(\data_\orig,\data_\hijack) = \decoder \Big (\encoder_\orig(\data_\orig) || \encoder_\hijack(\data_\hijack) \Big )      = \data_\camD,
\]
where $\data_\orig$ denotes a sample from the original dataset ($\data_\orig \in \dataset_\orig$), $\data_\hijack$ a sample from the hijacking dataset ($\data_\hijack \in \dataset_\hijack$), and $\data_\camD$ the camouflaged sample. 

\mypara{Visual Loss ($\loss_{vl}$)}
Next, we introduce the first loss of the \camM, i.e., the Visual Loss.
This loss drives the {\camM} to output data that has the visual appearance as samples in the hijackee dataset. 
Intuitively, the Visual Loss calculates the L1 distance between the output of the {\camM} and the hijackee sample.
More formally, we define the Visual Loss as follows:
\[
\loss_{vl} = \min\lVert \data_\camD -  \data_\orig \lVert,
\]

\mypara{Semantic Loss}
The Visual Loss associates the {\camM}'s output with the hijackee sample on the visual perspective.
We now introduce the Semantic Loss which associates the {\camM}'s output to the features of the hijacking sample.
Since the Semantic Loss operates on the feature level and not the visual level, we first need a feature extractor $\feat$ which extracts the features of a given sample.
This feature extractor $\feat$ can for example be a middle layer of any classification model.
Since we do not assume the adversary's knowledge of any information about the target model as previously mentioned in \autoref{sec:threatModel}, we use a pretrained MobileNetV2~\cite{SHZZC18} as our feature extractor.
However, a stronger adversary that has access to the target model can use the target model as the feature extractor.
The output of any layer of $\feat$ can be picked as the features. 
For our work, we use the output of the second to last layer of MobileNetV2.

Intuitively, the Semantic Loss calculates the L1 distance between the features of the output of the {\camM} and the hijacking sample.
More formally, we define the Semantic Loss as follows:
\[
\loss_{sl} = \min\lVert \feat(\data_\camD) -  \feat(\data_\hijack) \lVert,
\]
where  $\feat$ is the feature extractor.

\mypara{Adverse Semantic Loss}
So far the Visual and Semantic Losses already associate the {\camM}'s output with both the visual appearance of the hijackee sample and the features of the hijacking sample.
However, in certain cases when the hijacking and hijackee datasets are complex and similar, as shown later, the camouflaged sample's features are not distinct enough from the hijackee sample's features, which degrades the performance of the model hijacking attack.
Hence, we introduce another loss, i.e., the Adverse Semantic Loss.
This loss maximizes the difference between the features of the hijackee and camouflaged samples using the L1 distance.
We define the adverse Semantic Loss as:
\[
\loss_{asl} = \max\lVert \feat(\data_\camD) -  \feat(\data_\orig) \lVert,
\]

%-------------------------------------------------------------------------------
\subsection{The {\firstAttack} Attack}
\label{sec:AttackMeth}
%-------------------------------------------------------------------------------

After presenting the general pipeline and the buildings blocks, we now present our first concrete model hijacking attack, namely the {\firstAttack} attack.

Intuitively, the {\firstAttack} attack uses a {\camM} to camouflage the hijacking dataset and poison the target model. 
The {\firstAttack} attack can be divided into three stages, namely preparatory, camouflaging, and executing. 
We now explain each of them in detail.

\mypara{Preparatory}
In the first stage, the adversary setups their hijacking and hijackee datasets.
To recap, this hijackee dataset is used for camouflaging the hijacking dataset. 

Next, after creating the hijackee dataset, the adversary creates a mapping between the original dataset's labels and the ones in the hijacking dataset.
In this work, we assign the labels in a non-semantic manner. 
More concretely, we assign the $i^{th}$ label from the original dataset to the $i^{th}$ label of the hijacking dataset, irrespective of what each label stands for.
However, our attack is independent of the mapping technique and the adversary can freely create the mapping with the only restriction of keeping it consistent throughout the attack.

Finally, the adversary picks their feature extractor $\feat$ which is used to calculate the features of samples needed to train the {\camM}.
As previously mentioned (\autoref{sec:buildingBlocks}), the feature extractor is an off-the-shelf model that the adversary can freely choose.

\mypara{Camouflaging}
After deciding on the hijackee dataset, label mapping, and the feature extractor, the adversary can now start the main process of the {\firstAttack} attack.
We use both of the Visual and Semantic Losses to build the {\camM} for this attack.

As previously mentioned and demonstrated in~\autoref{fig:camMStuct}, the {\camM} uses two encoders and a single decoder.
All three models, i.e., the two encoders and the decoder, are trained jointly with both losses.
More formally we define the loss as the following.
\begin{equation}
\label{eq:Cham}
\lossM_{Cham}(\data_\camD, \data_\orig,\data_\hijack) = \min \Big ( \lVert \data_\camD -  \data_\orig  \lVert +\lVert \feat(\data_\camD) -  \feat(\data_\hijack) \lVert\Big )
\end{equation}
As the loss demonstrates, the {\camM} is independent of the target model.
Hence, it can be used to hijack multiple target models with a similar original task.

For the {\firstAttack} attack, the adversary uses the hijackee dataset and the hijacking dataset to train the {\camM}. 
More concretely, the adversary trains their {\camM} as follows:
\begin{enumerate}
\setlength\itemsep{1em}
\item For each epoch, the adversary first randomly pairs samples from the hijackee dataset to the samples in the hijacking dataset.
Since both datasets can be of different sizes, the mapping between both datasets can be many-to-many instead of one-to-one. 
We change the mapping in each epoch to increase the generalizability of the {\camM}.
\item After mapping the samples, we feed each pair (a hijackee sample and a hijacking sample) to the {\camM}.
Finally, the {\camM}'s output and the input samples are used to update the {\camM} using \autoref{eq:Cham} as the loss function.
\end{enumerate}

\begin{figure*}[!t]
\centering
\includegraphics[width=1.4\columnwidth]{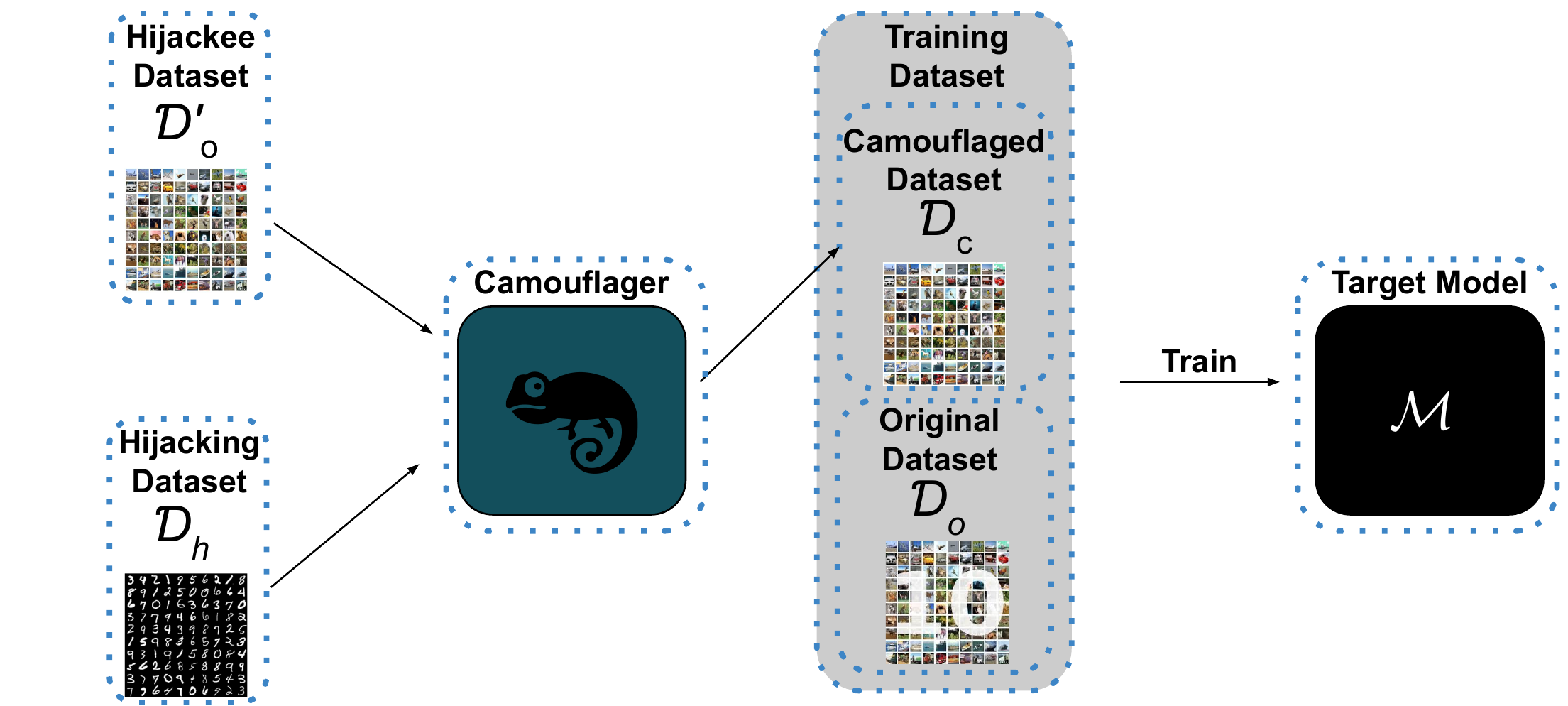}
\caption{An overview of the model hijacking attack. 
First, the adversary inputs the hijackee and hijacking datasets to the {\camM}. 
Next, they take the {\camM}'s output (the camouflaged dataset) and poison the training dataset of the target model. 
Finally, the model is trained with the poisoned dataset.}
\label{fig:attackOverview}
\end{figure*}

\mypara{Executing}
After training the {\camM}, the adversary can now execute their attack.
\autoref{fig:attackOverview} shows an overview of the {\firstAttack} attack after the training of {\camM}.
As the figure shows, first, the adversary maps the samples inside the hijackee dataset to samples from the hijacking dataset and creates the camouflaged dataset by querying the trained {\camM}.
To recap, the labels used to create the camouflaged dataset are the ones from the hijacking dataset.
Next, they use the camouflaged dataset to poison the training of the target model to hijack it.
We refer to the target model after being trained using the poisoned dataset as the \emph{hijacked model}.

After hijacking the target model, the adversary can query any sample from the hijacking dataset's distribution by first camouflaging it using the {\camM}.
Then, they query the camouflaged sample to the hijacked model, and map the predicted label to its corresponding label in the hijacking dataset.

%-------------------------------------------------------------------------------
\subsection{The {\secondAttack} Attack}
\label{sec:advAttackMeth}
%-------------------------------------------------------------------------------

As will be shown later in \autoref{sec:eval}, the {\firstAttack} attack has good performance when the hijacking and hijackee datasets are significantly different.
However, when both datasets are complex and not significantly different, the performance starts degrading.
Hence, we propose the more advanced attack, namely the {\secondAttack} attack.

The {\secondAttack} attack tries to explicitly distance the features of the output of the {\camM} from the hijackee dataset.
To accomplish this, in addition to the Visual and Semantic Losses, we use the adverse Semantic Loss.
More formally instead of using \autoref{eq:Cham} as the loss for training the {\camM}, we use the following loss.

\begin{equation}
\begin{split}
\label{eq:ChamAdv}
\lossM_{ChamAdv}(\data_\camD, \data_\orig,\data_\hijack) = \min \Big ( \lVert \data_\camD -  \data_\orig  \lVert +\lVert \feat(\data_\camD) -  \feat(\data_\hijack) \lVert \\- \lVert \feat(\data_\camD) -  \feat(\data_\orig) \lVert\Big )
\end{split}
\end{equation}
Besides the different loss function, to execute the {\secondAttack} attack, the adversary follows the same steps as the {\firstAttack} attack (\autoref{sec:advAttackMeth}).

%-------------------------------------------------------------------------------
\section{Evaluation}
\label{sec:eval}
%-------------------------------------------------------------------------------

In this section, we present our experimental results.
We start by introducing our datasets and evaluation settings.
Next, we evaluate our {\firstAttack} and {\secondAttack} attacks.
Finally, we study the impact of some of the hyperparameters in our model hijacking attacks.

%-------------------------------------------------------------------------------
\subsection{Datasets Description}
%-------------------------------------------------------------------------------

To evaluate our different model hijacking attacks, we use three  well-established computer vision benchmark datasets, namely MNIST, CIFAR-10, and CelebA.
We now briefly introduce them:

\mypara{MNIST}
MNIST is a grey-scale handwritten digits classification dataset.
It consists of 70,000 images, each of them is in the size of $28\times28$ and contains a single digit at its center.
The MNIST dataset is equally split between 10 classes.

Since the state-of-the-art machine learning models we use in our work, e.g., the MobileNetV2~\cite{SHZZC18} and Resnet18~\cite{HZRS16} models, expects inputs with the size of $224\times 224$, we rescale the MNIST dataset to satisfy it.
Moreover, we convert the grey-scale images to three channels images, by repeating the same values in all channels to be able to use the MNIST dataset on the same models trained on colored -- three channels -- images.

\mypara{CIFAR-10} 
CIFAR-10 is a 10 classes colored dataset.
It consists of 60,000 images with the size of $32 \times 32$.
The images are equally split between the following 10 classes: Airplane, automobile, bird, cat, deer, dog, frog, horse, ship, and truck.
Similar to the MNIST dataset, we rescale the CIFAR-10 images to $224\times 224$.

\mypara{CelebA} 
CelebA is a dataset of face attributes with more than 200,000 colored images.
We use the aligned version of the dataset, where each image contains the face of a celebrity in the middle of it and is labeled with 40 different binary attributes.
We follow Salem et al.~\cite{SWBMZ20} to create an 8-class classification task by concatenating the top three balanced attributes, i.e., Heavy Makeup, Mouth Slightly Open, and Smiling.
We randomly sample 40,000 images for training and 5,000 for testing.
Finally,it is important to note that unlike the MNIST and CIFAR-10 datasets, the CelebA dataset is highly imbalanced.

%-------------------------------------------------------------------------------
\subsection{Evaluation Settings}
%-------------------------------------------------------------------------------

We now introduce our evaluation settings.
We start with the model structures we use, then we present our evaluation metrics.

%-------------------------------------------------------------------------------
\subsubsection{Model Structures}
\label{sec:evalSettings}
%-------------------------------------------------------------------------------

As previously mentioned, for this work, we focus on the machine learning classification setting.
To this end, we use a state-of-the-art classification model for our target models (original task), namely Resnet18~\cite{HZRS16}.

Since we do not assume any knowledge about the target model for our model hijacking attacks, we use a completely different model as our feature extractor.
More concretely, we use MobileNetV2~\cite{SHZZC18}.

For the {\camM}, we use the following architecture for both of the encoders ($\encoder_\orig$ and $\encoder_\hijack$):
\begin{tcolorbox}[boxsep=1pt,left=2pt,right=2pt,top=0.5 pt,bottom=0pt]
\emph{The {\camM} encoders ($\encoder_\orig$ and $\encoder_\hijack$) architecture:}
\begin{align*}
\data_{in} \rightarrow
\texttt{Conv2d(4,12)} & \\
\texttt{Conv2d(4,24)} & \\
\texttt{Conv2d(4,48)} & \\
\texttt{Conv2d(4,96)}&
\\
\rightarrow \mu
\end{align*}
\end{tcolorbox}
\noindent Here, $\data_{in}$ is the input sample, \texttt{conv2d(k,f)} is a two dimensional convolution layer with kernel of size \texttt{k} and \texttt{f} filters, and $\mu$ is the encoder's output latent vector.
After each convolution layer, we apply batch normalization and adopt \texttt{ReLU} as the activation function.

Finally, for the {\camM}'s decoder we use the following architecture:
\begin{tcolorbox}[boxsep=1pt,left=2pt,right=2pt,top=0.5 pt,bottom=0pt]
\emph{The {\camM} decoder ($\decoder$) architecture:}
\begin{align*}
(\mu_\orig||\mu_\hijack) \rightarrow
\\
\texttt{ConvTranspose2d(4,96)} & \\
\texttt{ConvTranspose2d(4,48)} & \\
\texttt{ConvTranspose2d(4,24)} & \\
\texttt{ConvTranspose2d(4,3)}&
\\
\rightarrow \data_{out}
\end{align*}
\end{tcolorbox}
\noindent Here, $(\mu_\orig||\mu_\hijack)$ is the concatenation of the latent vectors for both the original and hijacking samples after being encoded with the $\encoder_\orig$ and $\encoder_\hijack$ encoders, respectively.
\texttt{ConvTranspose2d(k',f')} is a two dimensional transposed convolution layer with kernel of size \texttt{k'} and \texttt{f'} filters, and $\data_{out}$ is the output camouflaged sample.
After each layer, we apply batch normalization and adopt \texttt{ReLU} as the activation function, except for the last layer where we only use the \texttt{Tanh} activation function (to restrict the range for the decoded camouflaged sample).

Finally, we use the Adam optimizer to train the {\camM}.

%-------------------------------------------------------------------------------
\subsubsection{Evaluation Metrics}
%-------------------------------------------------------------------------------

To evaluate the performance of our model hijacking attack, we two metrics, namely \emph{Utility} and \emph{Attack Success Rate}.

\mypara{Utility}
Utility measures how close the performance of the hijacked model is to a clean, i.e., non-hijacked, model on the original dataset.
The closer the performance of the hijacked and clean models, the better the model hijacking attack.
More concretely, to measure the utility, we compare the accuracy of both the hijacked and clean models on a clean testing dataset, i.e., a testing dataset from the same distribution as the original dataset.

\mypara{Attack Success Rate}
The Attack Success Rate measures the model hijacked attack performance on the hijacking dataset.
We calculate the Attack Success Rate by computing the accuracy of the hijacked model on a hijacking testing dataset, i.e., a testing dataset with the same distribution as the hijacking dataset.
For both of our {\firstAttack} and {\secondAttack} attacks, we first camouflage the hijacking testing dataset before querying it to the hijacked model and calculate the accuracy.

\begin{figure}[!t]
\centering
\begin{subfigure}{0.49\columnwidth}
\includegraphics[width=\columnwidth]{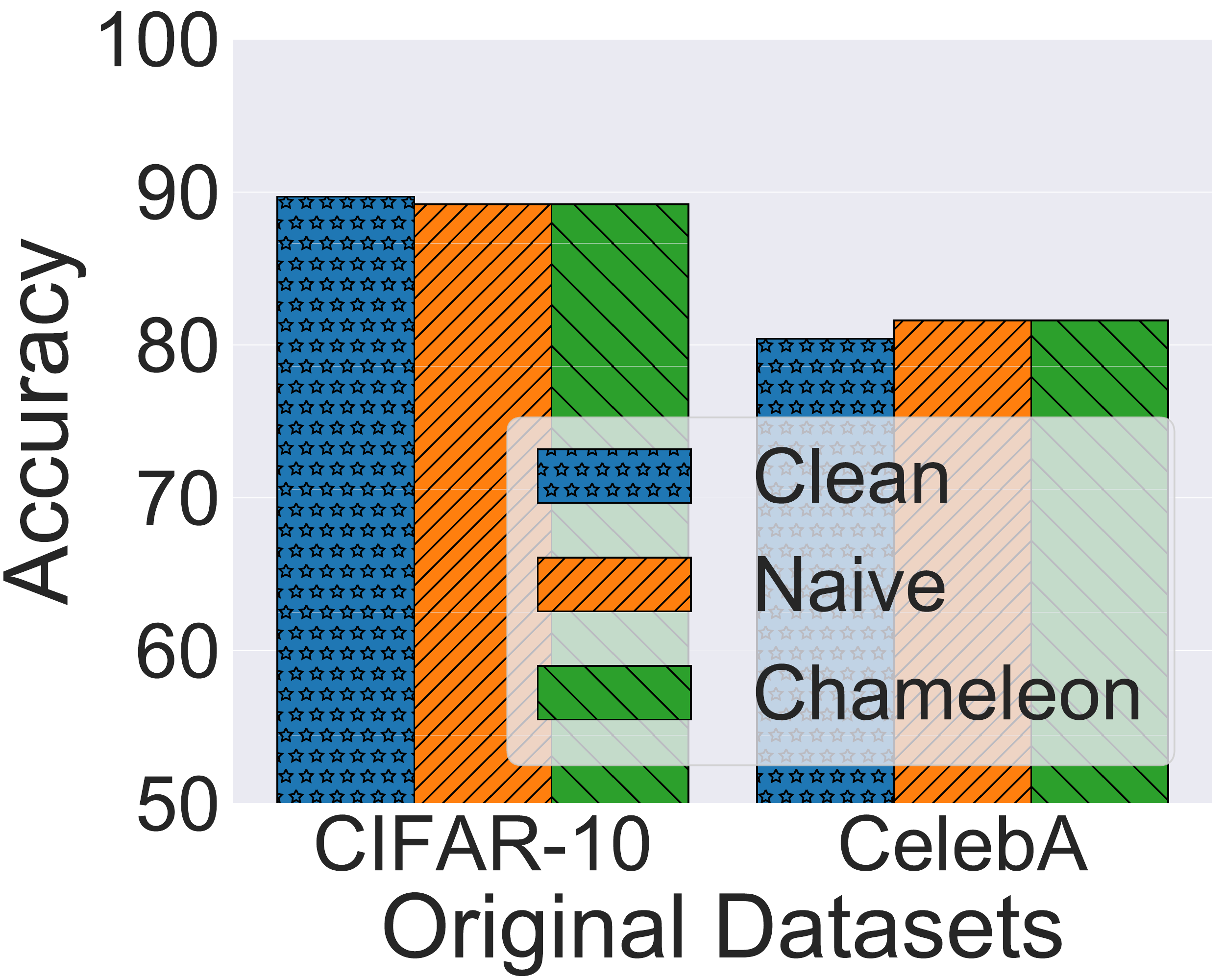}
\caption{Utility}
\label{fig:utlity1stAttk} 
\end{subfigure}
\begin{subfigure}{0.49\columnwidth}
\includegraphics[width=\columnwidth]{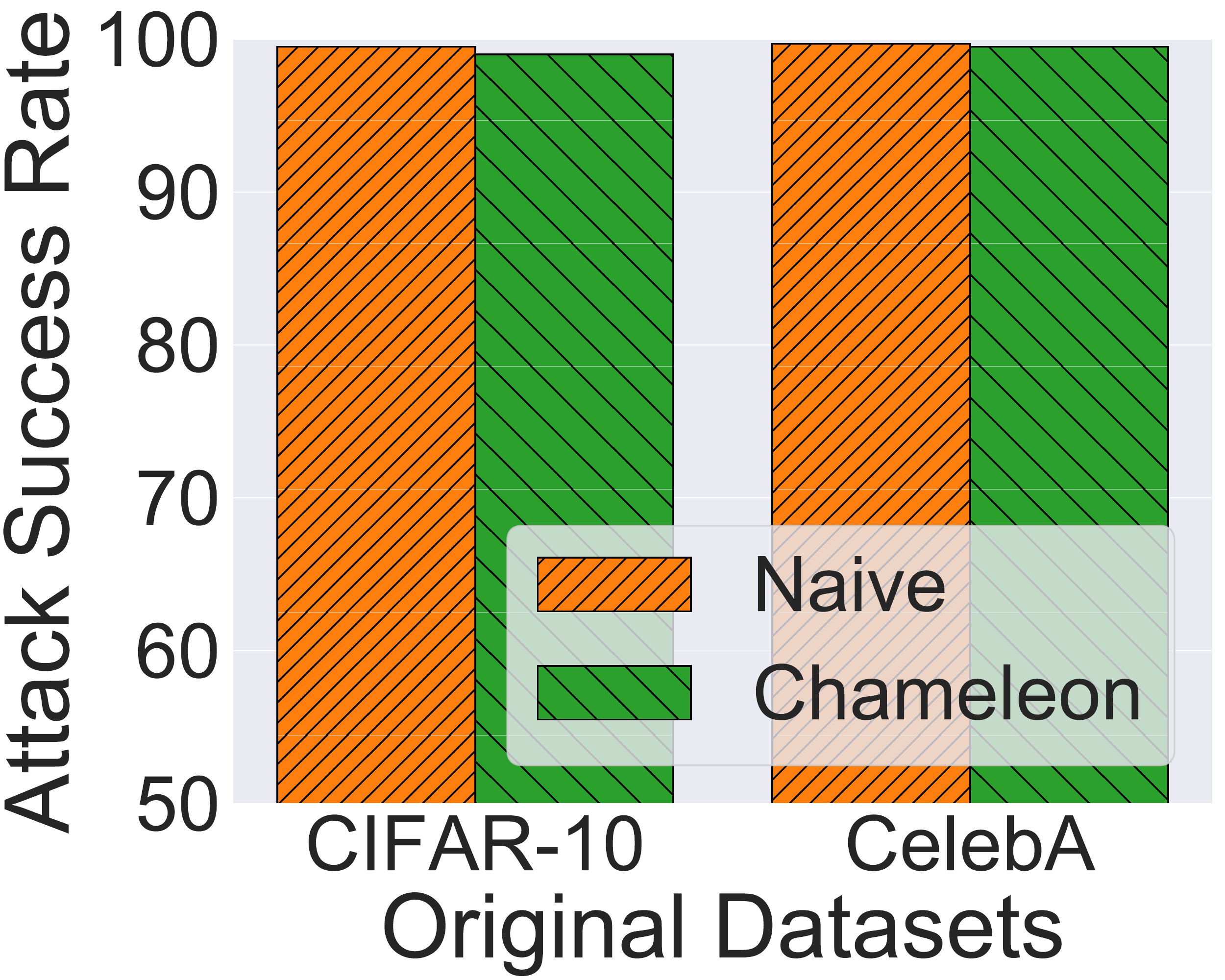}
\caption{Attack Success Rate}
\label{fig:ASR1stAttk} 
\end{subfigure}
\caption{The results of our {\firstAttack} Attack. 
The original datasets are noted on the x-axis.
For both (CIFAR-10 and CelebA) datasets, we use MNIST as the hijacking dataset.
\emph{Naive} corresponds to applying the model hijacking attack without camouflaging the hijacking dataset.
\autoref{fig:utlity1stAttk} compares the Utility of both the Naive and {\firstAttack} attacks with a clean model (Clean) using the original testing dataset, and \autoref{fig:ASR1stAttk} compares the Attack Success Rate of both attacks on the hijacking testing dataset.}
\label{fig:perf1stAttk}
\end{figure}

\begin{figure}[!t]
\centering
\begin{subfigure}{0.7\columnwidth}
\includegraphics[width=\columnwidth]{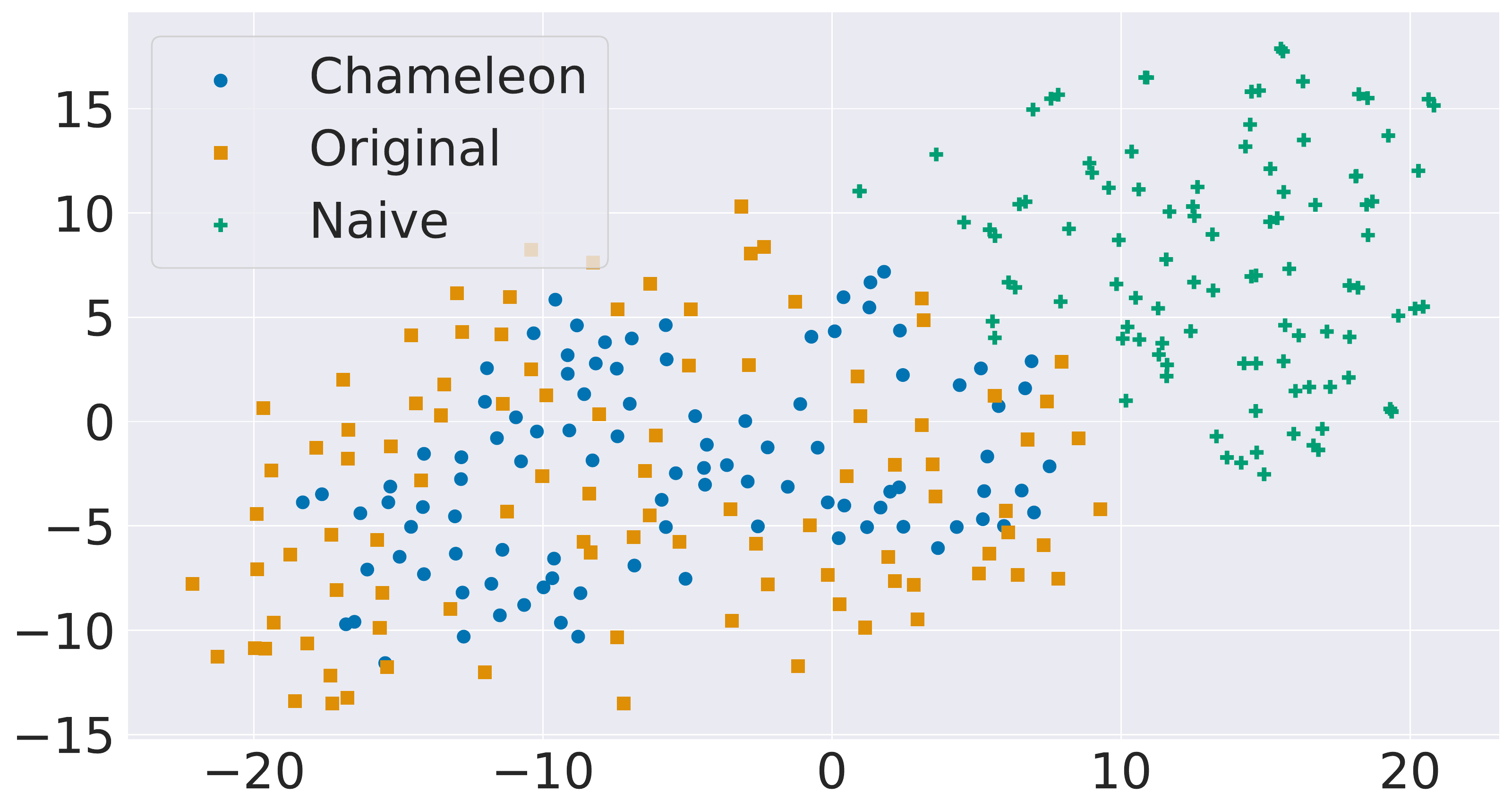}
\caption{CIFAR-10}
\label{fig:cifarTsne1stAttk} 
\end{subfigure}
\begin{subfigure}{0.7\columnwidth}
\includegraphics[width=\columnwidth]{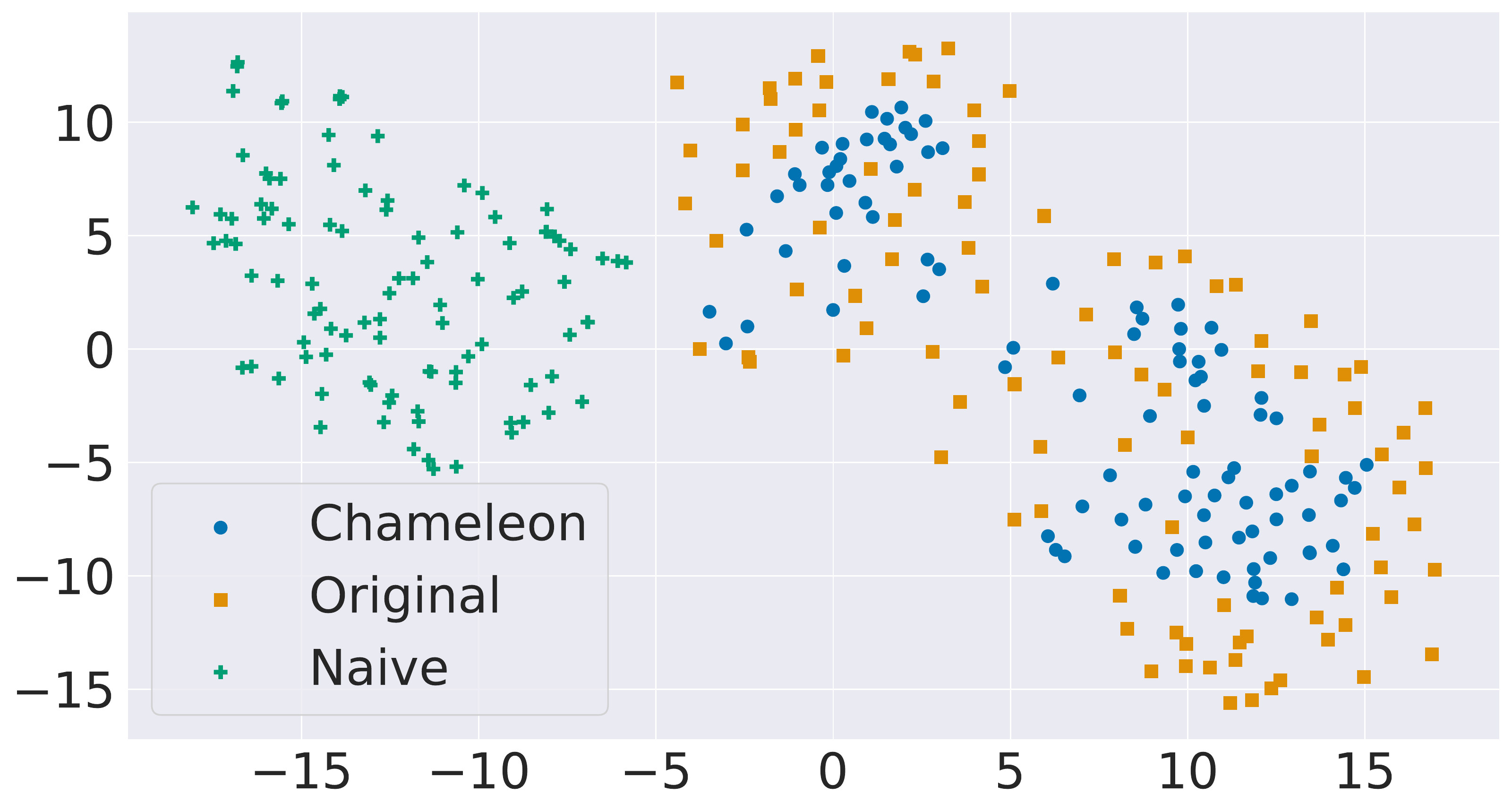}
\caption{CelebA}
\label{fig:celebATsne1stAttk} 
\end{subfigure}
\caption{ Visualization of the difference in stealthiness between the {\firstAttack} and Naive attacks.
We use t-SNE to reduce the camouflaged, original, and hijacking samples to two dimensions.
Then, we plot them in \autoref{fig:cifarTsne1stAttk} for the CIFAR-10 dataset, and \autoref{fig:celebATsne1stAttk} for the CelebA dataset.
Here MNIST is the hijacking dataset, and CIFAR-10 (\autoref{fig:cifarTsne1stAttk}) and CelebA (\autoref{fig:celebATsne1stAttk}) are the original datasets.}
\label{fig:tsne1stAttk}
\end{figure}

\begin{figure*}[!t]
\centering
\begin{subfigure}{0.65\columnwidth}
\includegraphics[width=\columnwidth]{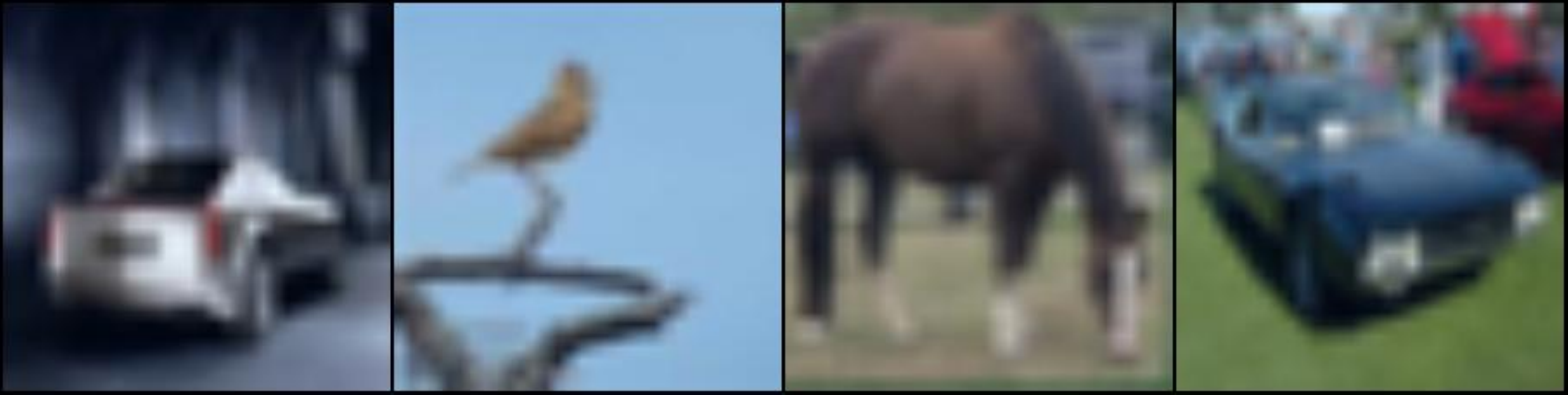}
\caption{Original CIFAR-10}
\label{fig:CIFAROrig} 
\end{subfigure}
\begin{subfigure}{0.65\columnwidth}
\includegraphics[width=\columnwidth]{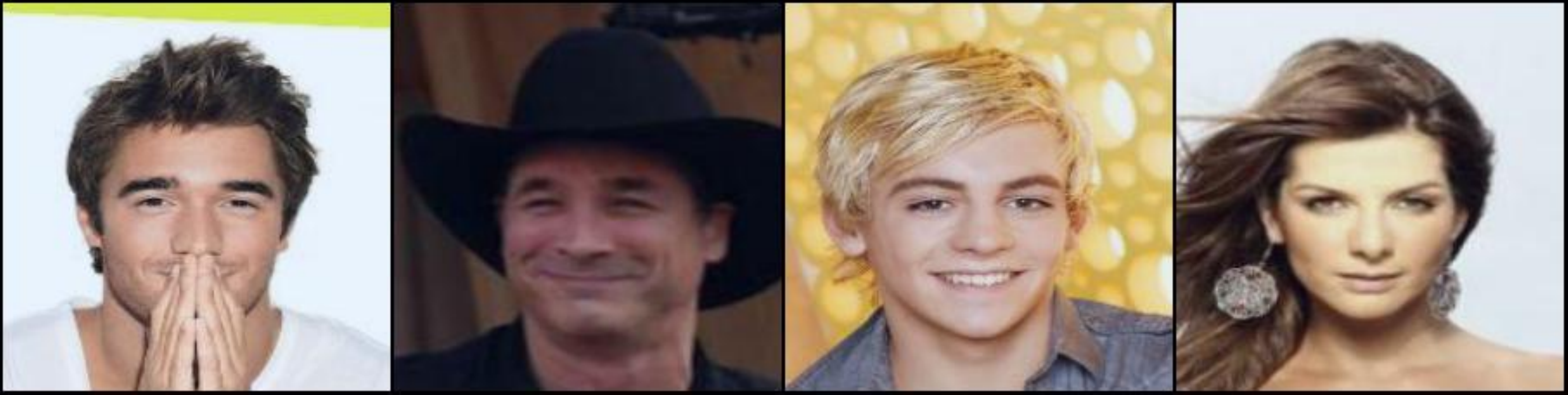}
\caption{Original CelebA}
\label{fig:CelebAOrig} 
\end{subfigure}
\begin{subfigure}{0.65\columnwidth}
\includegraphics[width=\columnwidth]{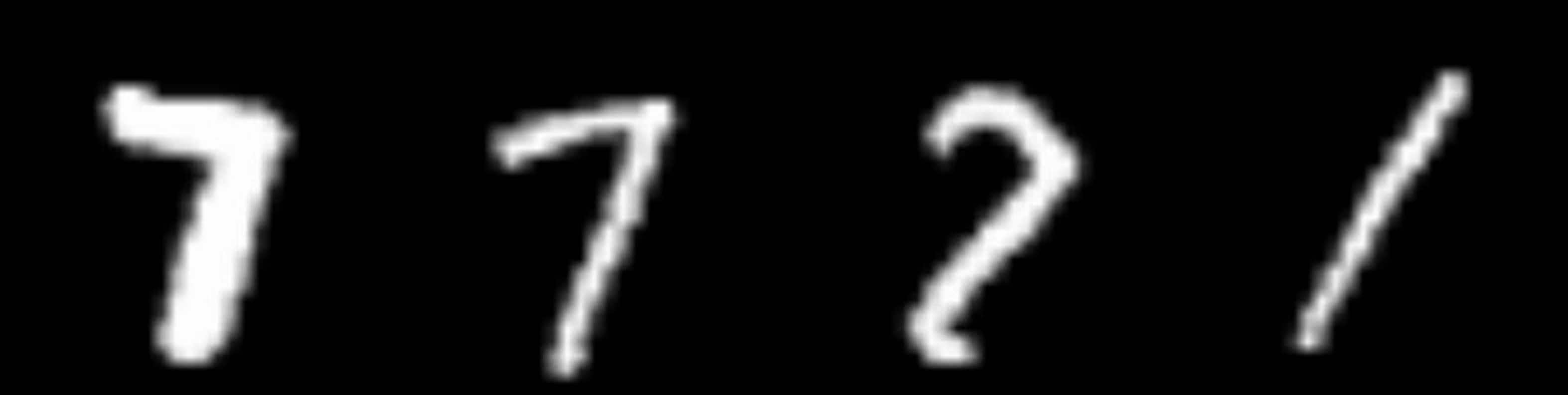}
\caption{Original MNIST}
\label{fig:MNISTOrig} 
\end{subfigure}
\begin{subfigure}{0.65\columnwidth}
\includegraphics[width=\columnwidth]{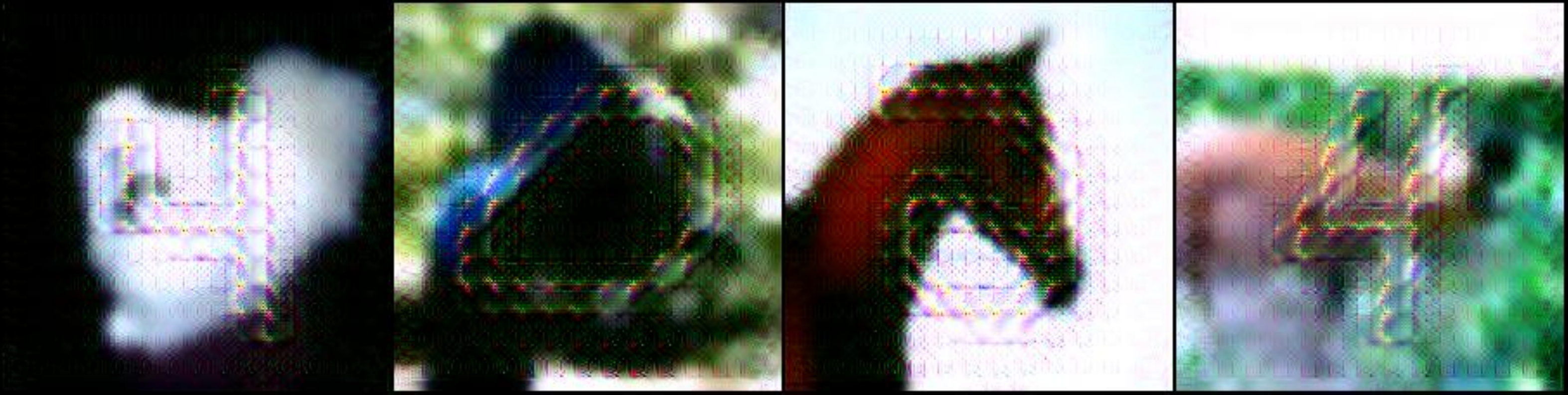}
\caption{Camouflaged CIFAR-10}
\label{fig:1stAttkCIFARVis} 
\end{subfigure}
\begin{subfigure}{0.65\columnwidth}
\includegraphics[width=\columnwidth]{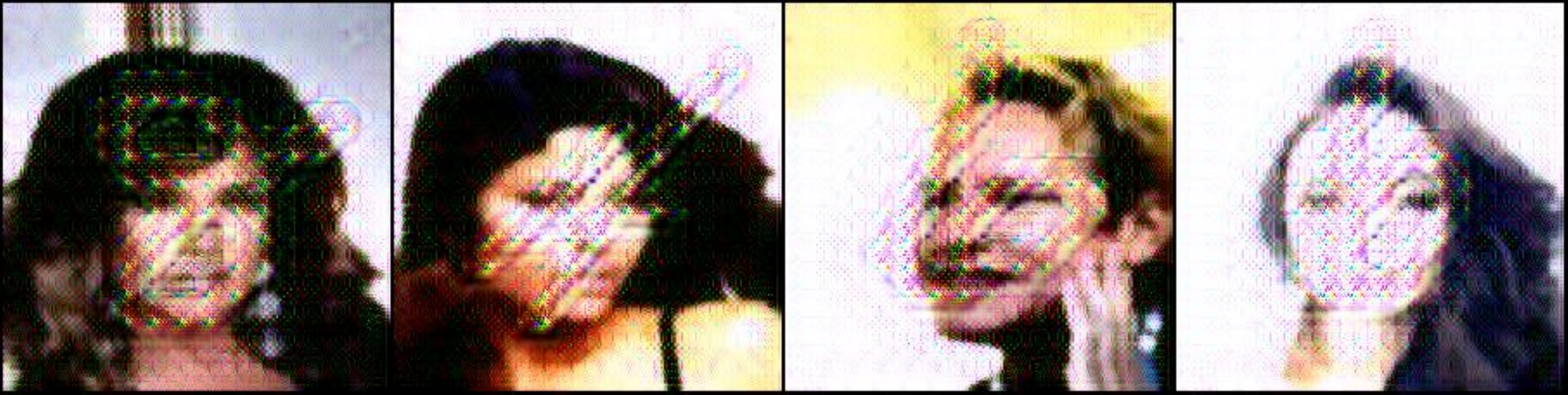}
\caption{Camouflaged CelebA}
\label{fig:1stAttkCelebAVis} 
\end{subfigure}
\caption{Visualization of the output of the {\camM} for the {\firstAttack} Attack for both the CIFAR-10 (\autoref{fig:1stAttkCIFARVis}) and CelebA (\autoref{fig:1stAttkCelebAVis}) datasets. 
Moreover, we show samples for both the Original(\autoref{fig:CIFAROrig} and \autoref{fig:CelebAOrig}) and hijacking (\autoref{fig:MNISTOrig}) datasets.}
\label{fig:1stAttkVis}
\end{figure*}

%-------------------------------------------------------------------------------
\subsection{The {\firstAttack} Attack}
\label{sec:1stAttkeval}
%-------------------------------------------------------------------------------

After introducing the datasets and our evaluations metrics, we now evaluate the performance of our {\firstAttack} attack.
We use MNIST as our hijacking dataset and both CIFAR-10 and CelebA as the original datasets for this attack.

Firstly, we map the labels of the hijacking dataset and the original dataset as mentioned in~\autoref{sec:AttackMeth}.
Next, we train the {\camM} by constructing two encoders and a decoder with the architecture presented in~\autoref{sec:evalSettings}, and a hijackee dataset by randomly sampling $1,000$ sample for $8$ random classes from the original dataset.
Then we randomly sample $10,000$ samples from the hijacking dataset and follow the methodology previously presented in~\autoref{sec:buildingBlocks} to train the {\camM}.

After training the {\camM}, we use it together with the same hijackee dataset to camouflage $40,000$ randomly sampled samples from the hijacking dataset.
Finally, we use the $40,000$ camouflaged samples together with their mapped labels to hijack the target model, i.e., we train the target model with both the camouflaged and original samples.

After presenting the concrete setup of our evaluation, we first evaluate the performance of the {\firstAttack} attack (\autoref{req:1} and \autoref{req:3}), then its stealthiness (\autoref{req:2} and \autoref{req:4}).

\mypara{Performance Evaluation}
To evaluate the performance of our {\firstAttack} attack, we consider the following baselines:
\begin{enumerate}
\setlength\itemsep{1em}
\item First, we train a clean model (Clean) using only the original dataset to compute and compare the Utility of the {\firstAttack} hijacked models.
\item Second, we perform the naive model hijacking attack (Naive),
where the adversary hijacks the target model without camouflaging the hijacking dataset first.
It is important to note that this naive attack serves as the upper bound of the Attack Success Rate performance, since the hijacking samples are used as is without any modifications to make them less stealthy, which is the goal of our advanced model hijacking attacks.
\end{enumerate}

We first compare the utility of our {\firstAttack} attack in~\autoref{fig:utlity1stAttk}.
As the figure shows, our {\firstAttack} attack achieves almost the same Utility as both the Clean and Naive models.
To recap, for reconstructing the Naive model we poison its training dataset with the hijacking datasets itself and not the camouflaged version.
More concretely, our {\firstAttack} attack achieves $89.2\%$ accuracy on the original testing dataset, which is exactly the same as the Naive attack and only $0.5\%$ lower than the Clean model for the CIFAR-10 dataset.
For the CelebA dataset, our {\firstAttack} and Naive hijacked models achieve $81.6\%$ accuracy which is $1.2\%$ better than the Clean model.
We believe this improved performance is due to the regularization effect of the extra poison data. 
Similar improvement of the CelebA classification models after data poisoning has been previously observed in backdoor attacks~\cite{SWBMZ20}.

Next, we compare the Attack Success Rate of our {\firstAttack} attack.
As the hijacking task here is MNIST, the Attack Success Rate measures the accuracy of the hijacking -- MNIST -- testing dataset.
As \autoref{fig:ASR1stAttk} shows, our {\firstAttack} attack achieves almost the same performance as the Naive hijacked models.
The {\firstAttack} hijacked model achieves $99\%$ Attack Success Rate when the original task is CIFAR-10 classification, which is only $0.5\%$ lower than the Attack Success Rate of the Naive hijacked model.
For the CelebA classification model, our attack achieves a $99.5\%$ Attack Success Rate, which is only $0.2\%$ lower than the one of the Naive model.

\begin{figure}[!t]
\centering
\includegraphics[width=0.7\columnwidth]{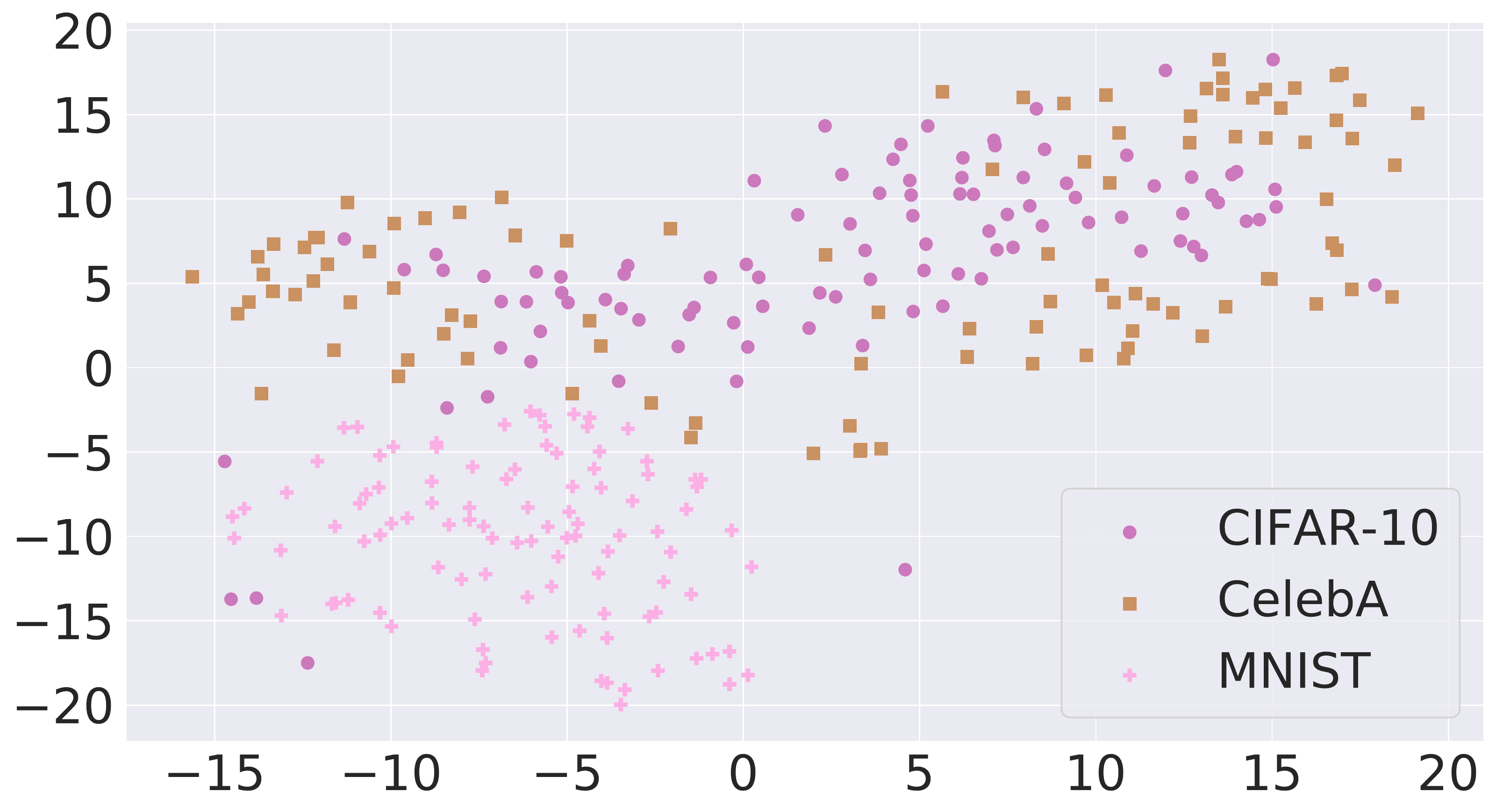}
\caption{Visualization of 100 random samples from the three datasets MNIST, CIFAR-10, and CelebA after reducing their dimension to two.
As the figure shows, the CIFAR-10 and CelebA datasets are clustered together, while MNIST can be separated from them. 
This shows the hardness of using one of the CelebA or CIFAR-10 datasets to hijack the other, unlike using the MNIST dataset.}
\label{fig:allDatasetstSNE}
\end{figure}

In general, hijacking models with the {\firstAttack} attack can achieve almost perfect Attack Success Rate (\autoref{req:3}) with a negligible drop in utility (\autoref{req:1}), which shows the efficacy of this attack when the original and hijacking datasets are significantly different (as will be shown later in \autoref{fig:allDatasetstSNE}).

\mypara{Stealthiness Evaluation}
Since two of the main requirements of our model hijacking attack focus on stealthiness (\autoref{req:2} and \autoref{req:4}), we now compare the stealthiness of the {\firstAttack} attack with the one of the Naive attack.
We use the following two approaches to measure the stealthiness:
\begin{enumerate}
\setlength\itemsep{1em}
\item First, we measure the Euclidean distance between the hijacking and the original datasets, as well as the camouflaged and original datasets.
To measure the Euclidean distance, we randomly sample 1,000 camouflaged, original, and hijacking samples.
Then for each camouflaged/hijacking sample, we find the closest original sample to it,
and calculate the Euclidean distance between them. 
We operate in a batch of 100 due to physical memory limitation.
Finally, we average the Euclidean distances of the camouflaged and hijacking samples independently.
\item Second, we use the t-distributed stochastic neighbor embedding (t-SNE)~\cite{MH08} for reducing 100 samples from the hijacking, original, and camouflaged datasets to two dimensions.
Then, we plot the reduced features of the samples.
\end{enumerate}

First for the Euclidean distance, our experiments show that our {\firstAttack} attack achieves $0.51$ Euclidean distance when hijacking a CIFAR-10 classification model using the MNIST hijacking dataset, which is about $82\%$ less than the one for the Naive attack ($0.93$).
Similarly, for the CelebA classification model, our {\firstAttack} attack achieves $0.77$ Euclidean distance, which is about $56\%$ less than the one of the Naive attack ($1.2$).
A lower distance denotes a more stealthy attack, since it shows that the two datasets are more similar.

Second, we visualize the t-SNE reduced samples for the CIFAR-10 and CelebA hijacked models in \autoref{fig:cifarTsne1stAttk} and \autoref{fig:celebATsne1stAttk}, respectively. 
As \autoref{fig:tsne1stAttk} clearly shows, the camouflaged ({\firstAttack}) samples are closer and hidden inside the original (Original) samples, unlike the hijacking (Naive) samples.
Note that in the Naive model hijacking attack, the adversary poisons the training dataset of the target model using the hijacking dataset itself.

As both the Euclidean distance and the visualization in \autoref{fig:tsne1stAttk} demonstrate, our {\firstAttack} attack distinctly outperforms the Naive model hijacking attack in terms of stealthiness (\autoref{req:2}).

Recall that our model hijacking attack has one more requirement with respect to the stealthiness of the attack (\autoref{req:4}), i.e., predicting samples from the hijacking dataset randomly if they are not camouflaged.
To this end, we run one more experiment to evaluate our {\firstAttack} hijacked models using the hijacking testing dataset without camouflaging. 
Our results show that indeed the accuracy on the non-camouflaged testing dataset is around $10\%$ for both cases, i.e., when using CIFAR-10 or CelebA as original datasets, which is the same as random guessing for the MNIST dataset

Finally, we visualize randomly sampled camouflaged samples together with ones from the original and hijacking datasets in ~\autoref{fig:1stAttkVis}.
Comparing figures \autoref{fig:1stAttkCIFARVis} and \autoref{fig:1stAttkCelebAVis} with figure \autoref{fig:CIFAROrig} and \autoref{fig:CelebAOrig}, we observe that indeed our camouflaged samples look like the original samples with some added artifacts.
Moreover, it is clear that the camouflaged samples (\autoref{fig:1stAttkCIFARVis} and \autoref{fig:1stAttkCelebAVis}) are visually more similar than the hijacking samples (\autoref{fig:MNISTOrig}), when compared to the original samples (\autoref{fig:CIFAROrig} and \autoref{fig:CelebAOrig}).

\begin{figure}[!t]
\centering
\begin{subfigure}{0.49\columnwidth}
\includegraphics[width=\columnwidth]{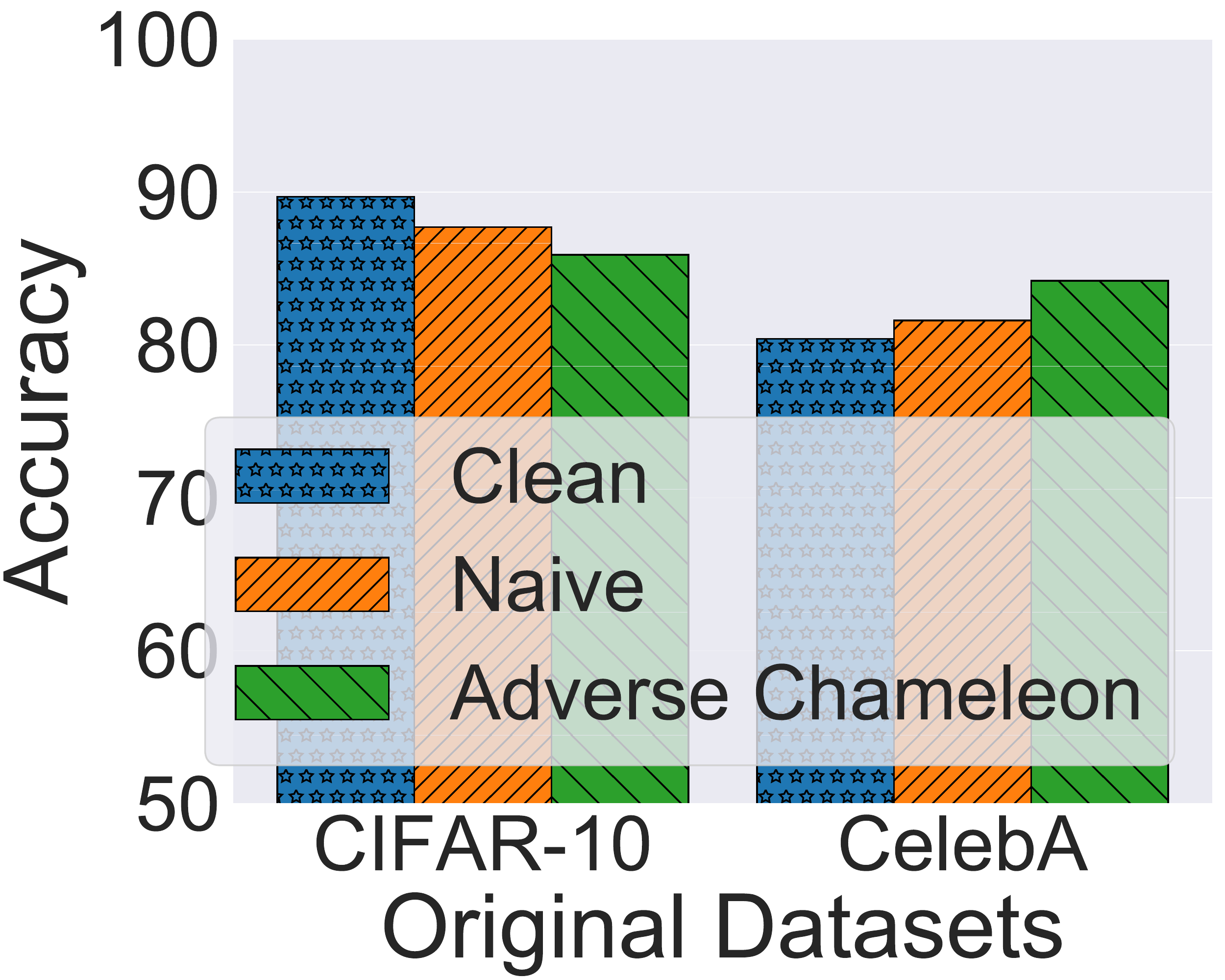}
\caption{Utility}
\label{fig:utlity2ndAttk} 
\end{subfigure}
\begin{subfigure}{0.49\columnwidth}
\includegraphics[width=\columnwidth]{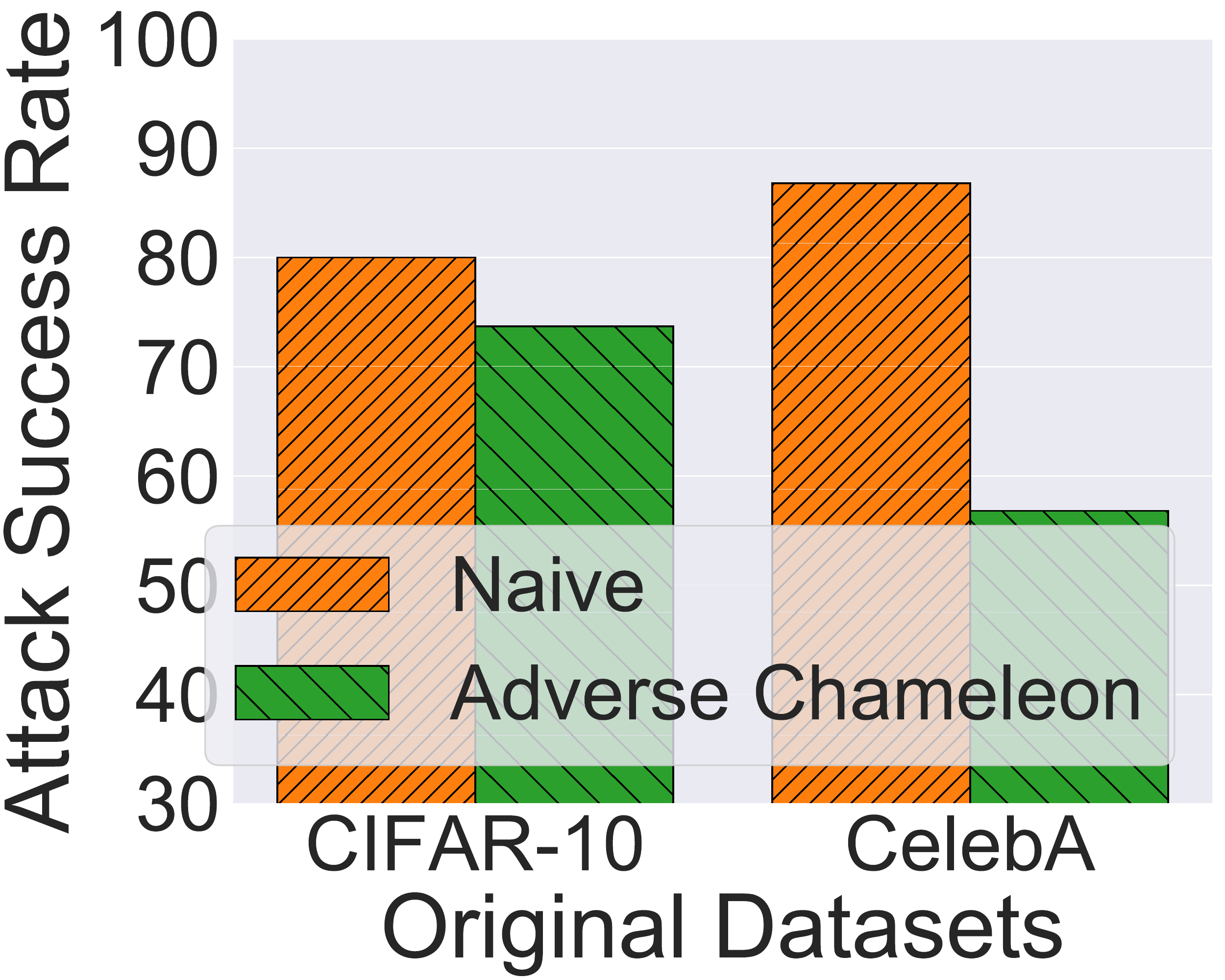}
\caption{Attack Success Rate}
\label{fig:ASR2ndAttk} 
\end{subfigure}
\caption{The results of our {\secondAttack} Attack. 
The original datasets are denoted on the x-axis, the hijacking dataset is CelebA when the original dataset is CIFAR-10 and vice versa.
Naive corresponds to applying the model hijacking attack without camouflaging the hijacking dataset first.
\autoref{fig:utlity2ndAttk} compares the utility of both the Naive and {\secondAttack} attacks with a clean model using the original testing dataset, and \autoref{fig:ASR2ndAttk} compares the Attack Success Rate of both attacks on the hijacking testing dataset.}
\label{fig:perf2ndAttk}
\end{figure}

%-------------------------------------------------------------------------------
\subsection{The {\secondAttack} Attack}
\label{sec:2ndAttkeval}
%-------------------------------------------------------------------------------

As previously shown (\autoref{sec:1stAttkeval}), our {\firstAttack} attack achieves strong performance when the hijacking and original datasets are significantly different.
However, when performing the {\firstAttack} attack using CIFAR-10 as the original dataset, and CelebA as the hijacking dataset, it only achieves an Attack Success Rate of $65.7\%$ which is $14.3\%$ less than the Naive attack.
We believe this gap between the two attacks is due to the following two reasons:
\begin{enumerate}
\setlength\itemsep{1em}
\item The first one is related to the more complex nature of the CelebA dataset compared to MNIST.
This can be seen in~\autoref{fig:1stAttkVis}, as the human faces have more information than the grey-scale digits.
\item Second, the CelebA and CIFAR-10 datasets are closer to each other compared to the MNIST dataset and either one of them.
To visualize this, we randomly sample 100 samples from each dataset and reduce each of the samples using t-SNE to two dimensions.
Then, we plot the results in \autoref{fig:allDatasetstSNE}.
As the figure shows, the CelebA and CIFAR-10 datasets are clustered together and can be separated from the MNIST dataset.
\end{enumerate}

Hence, when considering either of the CelebA or the CIFAR-10 datasets as the hijacking datasets, we execute the {\secondAttack} attack.
To evaluate the {\secondAttack} attack, we follow the same evaluation settings previously introduced in~\autoref{sec:1stAttkeval} with the following exception: Instead of using the {\firstAttack} attack to train the {\camM}, we use the {\secondAttack} attack to train it.
To recap, the {\secondAttack} attack uses the additional Adverse Semantic Loss to train the {\camM} together with both of the Visual and Semantic Losses.

After introducing the concrete setup of the {\secondAttack} attack we first evaluate its performance, then its stealthiness, and finally, we briefly discuss both the {\firstAttack} and {\secondAttack} attacks.

\mypara{Performance Evaluation}
We first compare the utility of our {\secondAttack} attack in \autoref{fig:utlity2ndAttk}.
As the figure shows, when using the {\secondAttack} attack to camouflage the CelebA dataset and hijack a CIFAR-10 classification model, the utility is only slightly dropped.
More concretely, the hijacked models achieve $87.7\%$ and $85.9\%$ accuracy on the CIFAR-10 testing dataset, when hijacking the models using the Naive and {\secondAttack} attacks, respectively.
This accuracy is only $2\%$, and $3.8\%$ less than the one of a clean CIFAR-10 classification model.
For the opposite case, i.e., the hijacking dataset is CIFAR-10 and the original dataset is CelebA, our {\secondAttack} attack achieves $84.2\%$ accuracy which is $2.6\%$ and $3.8\%$ higher than the one of the Naive attack and clean model, respectively.
We believe this increase in performance is due to the regularization effect of our attack.

Next, we evaluate the Attack Success Rate of our {\secondAttack} attack in~\autoref{fig:ASR2ndAttk}.
Specifically, we calculate the Attack Success Rate as the accuracy of the hijacking testing dataset when evaluating the Naive attack, and the camouflaged hijacking testing dataset when evaluating the {\secondAttack} attack.
As the figure shows, the Naive and {\secondAttack} attacks achieve
$80.0\%$ and $73.7\%$ Attack Success Rate when hijacking a CelebA classification model using the CIFAR-10 classification as the hijacking task, respectively.
For the other case, when the adversary aims to hijack a CIFAR-10 classification model to perform CelebA classification task, our {\secondAttack} attack achieves $56.8\%$, which is less than the one of Naive attack ($86.8\%$) but still significantly higher than random guessing.
Note that for this case, our {\secondAttack} attack achieves $2.6\%$ better utility than the Naive attack.

As both \autoref{fig:utlity2ndAttk} and \autoref{fig:ASR2ndAttk} show, our {\secondAttack} attack satisfies both of our performance related requirements, i.e., \autoref{req:1} and \autoref{req:3}.

\mypara{Stealthiness Evaluation}
Similarly to the {\firstAttack} attack, we evaluate the {\secondAttack} hijacked model against the non-camouflaged hijacking testing dataset.
As expected, the {\secondAttack} hijacked models achieve nearly random performance for the non-camouflaged hijacking testing dataset (\autoref{req:4}).
For instance, when using the CelebA dataset to hijack CIFAR-10, the non-camouflaged hijacking testing dataset achieves less than $20\%$ accuracy.

\begin{figure}[!t]
\centering
\begin{subfigure}{0.7\columnwidth}
\includegraphics[width=\columnwidth]{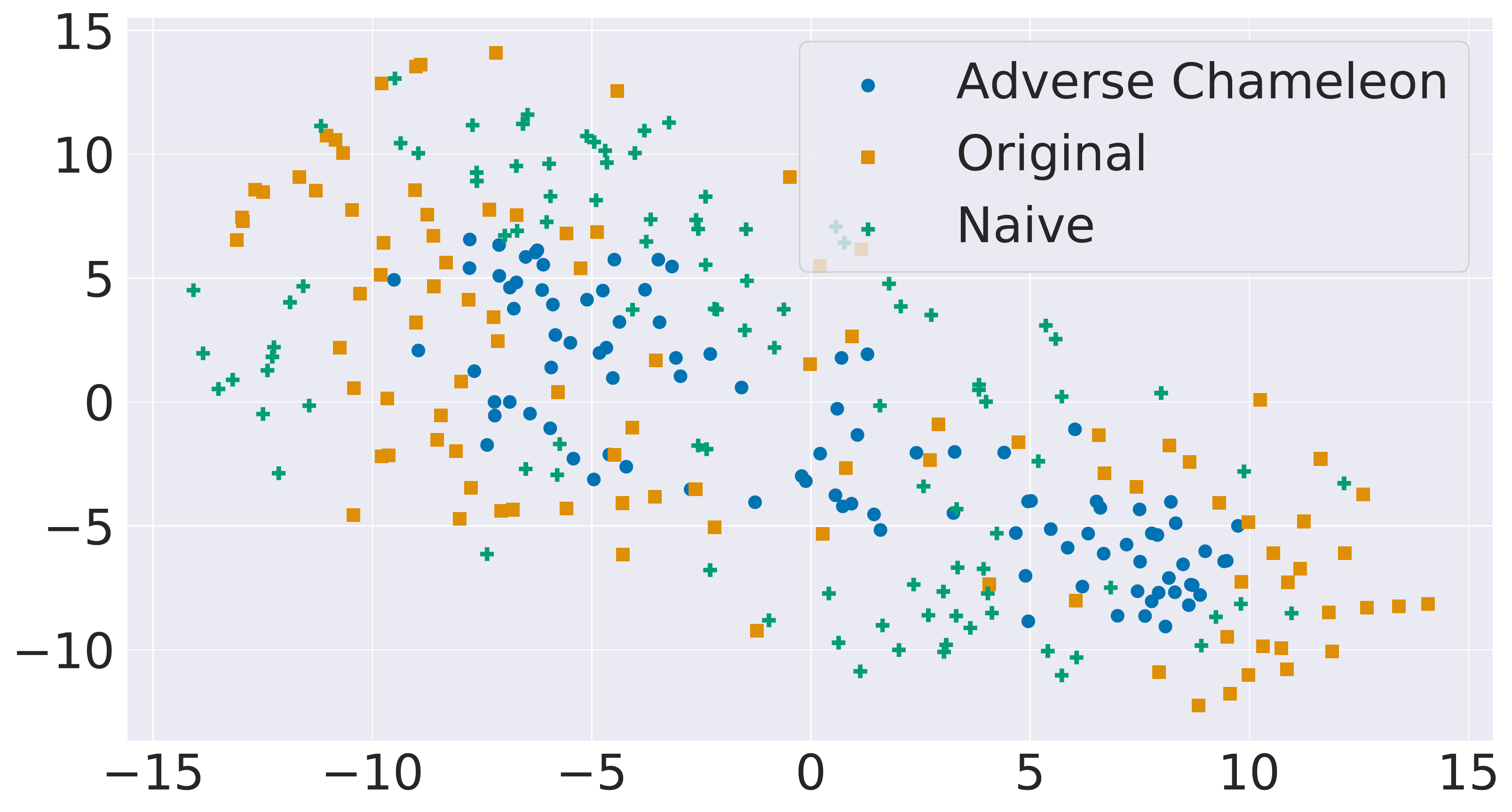}
\caption{CIFAR-10}
\label{fig:CIFARToCelebA_tSNE} 
\end{subfigure}
\begin{subfigure}{0.7\columnwidth}
\includegraphics[width=\columnwidth]{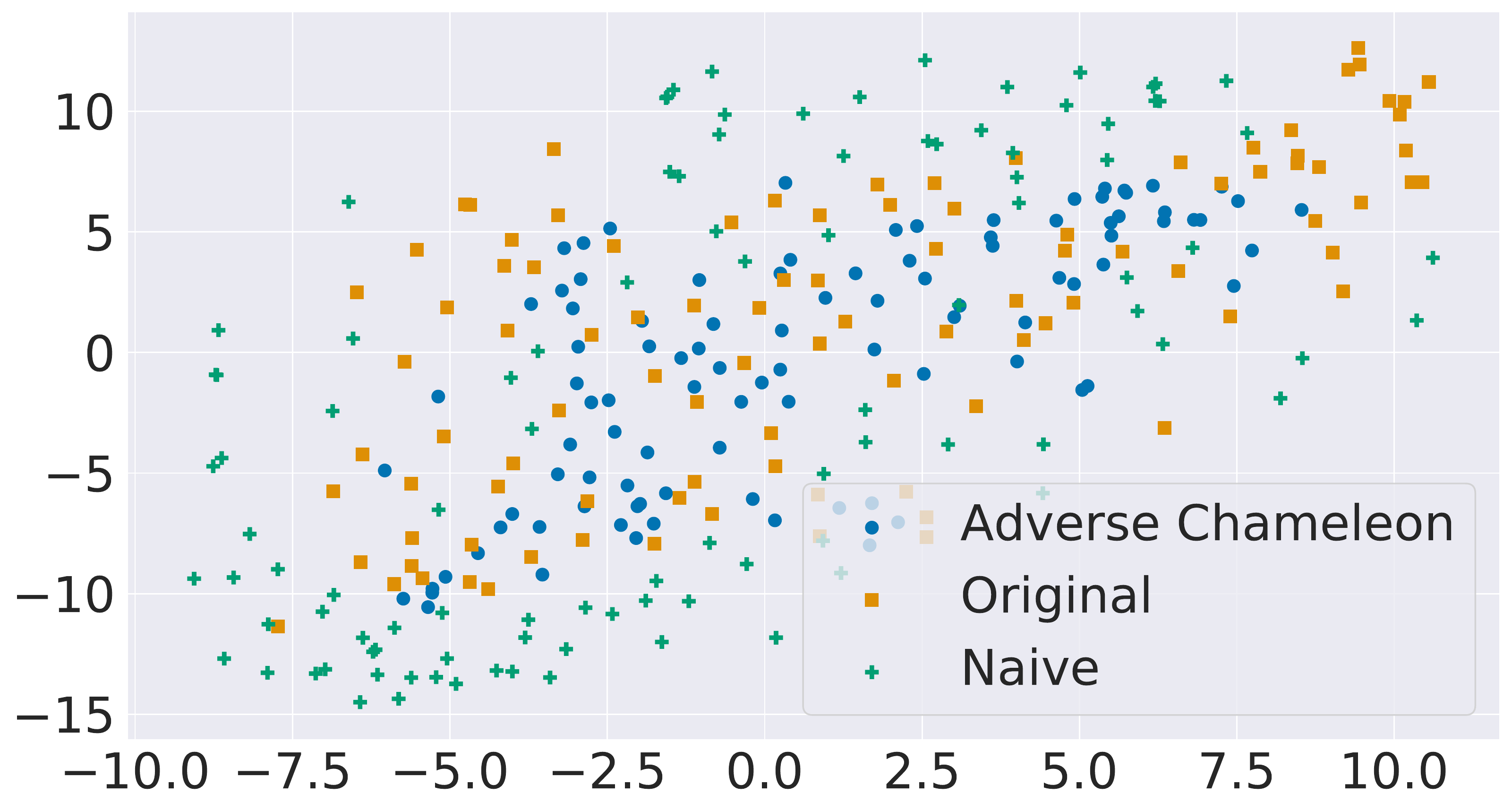}
\caption{CelebA}
\label{fig:CelebAtoCIFAR_tSNE} 
\end{subfigure}
\caption{ Visualization of the difference in stealthiness between {\secondAttack} and Naive attacks. We use t-SNE to reduce 100 camouflaged, original, and hijacking samples.
\autoref{fig:CIFARToCelebA_tSNE} shows the result when using CIFAR-10 as the hijacking dataset and CelebA the original dataset, and \autoref{fig:CelebAtoCIFAR_tSNE} shows the opposite case.}
\label{fig:tsne2ndAttk}
\end{figure}

\begin{figure}[!t]
\centering
\begin{subfigure}{\columnwidth}
\centering
\includegraphics[width=0.7\columnwidth]{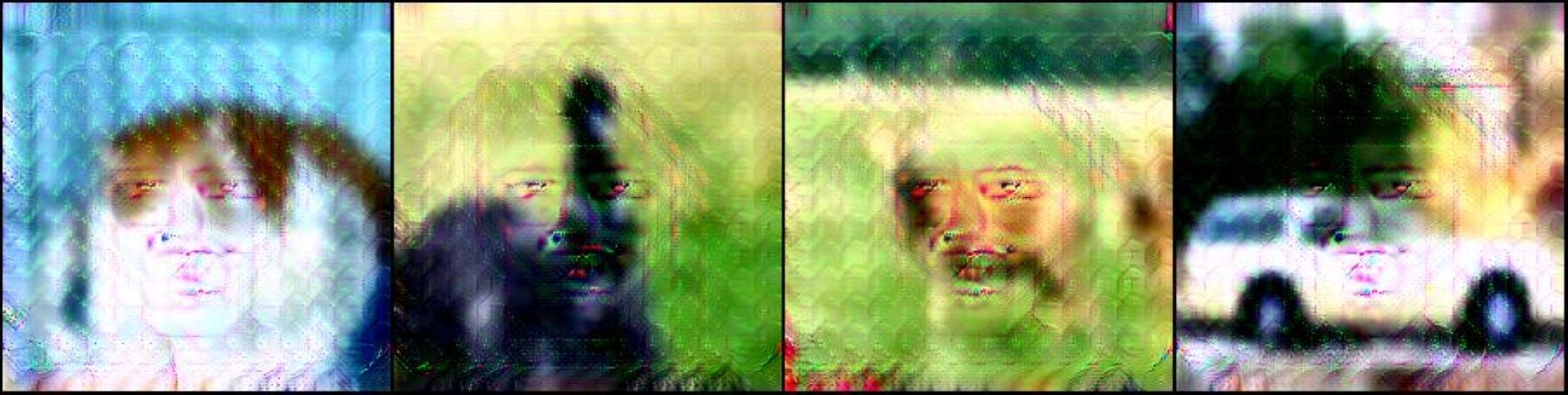}
\caption{Camouflaged CelebA}
\label{fig:2ndAtkCamoCelebA} 
\end{subfigure}
\begin{subfigure}{\columnwidth}
\centering
\includegraphics[width=0.7\columnwidth]{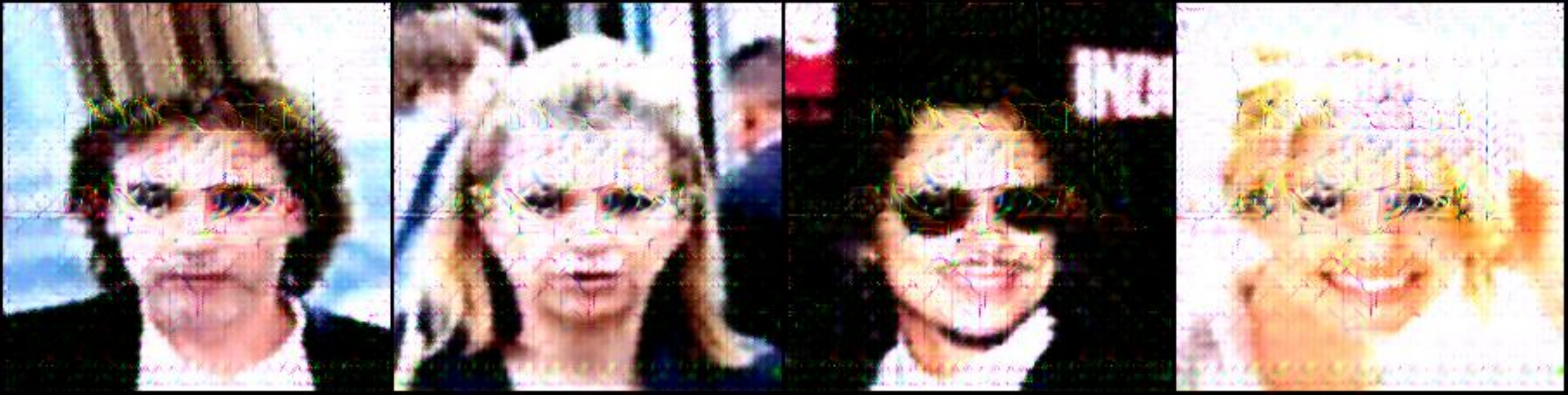}
\caption{Camouflaged CIFAR-10}
\label{fig:2ndAtkCamoCIFAR} 
\end{subfigure}
\caption{Visualization of the output of the {\camM} for the {\secondAttack} Attack.
We show the results when using the CelebA as the hijacking dataset and CIFAR-10 as the original dataset in \autoref{fig:CIFARToCelebA_tSNE}, and vice versa in\autoref{fig:CelebAtoCIFAR_tSNE}.}
\label{fig:2ndAttkVis}
\end{figure}

Next, we follow the same steps previously introduced in~\autoref{sec:1stAttkeval}
to visualize and compare the stealthiness (\autoref{req:2}) of the {\secondAttack} using the Euclidean distance and t-SNE.

First, we compare the Euclidean distance.
Using our {\secondAttack} attack to hijack a CIFAR-10 classification model with a CelebA classification -- hijacking -- task results in $0.52$ Euclidean distance.
This is less than the Euclidean distance of the Naive attack by a factor of $2.5$.
Similarly, when using the CelebA classification task as the original task and the CIFAR-10 classification as the hijacking task, our {\secondAttack} achieves $0.77$ Euclidean distance, which is $1.6$ times lower than the one with the Naive attack.

Second, we visualize the t-SNE reduced samples for all the original, hijacking, and camouflaged samples in ~\autoref{fig:tsne2ndAttk}.
In \autoref{fig:tsne2ndAttk}, the Naive samples directly correspond to the hijacking samples, since  to perform the Naive attack, the adversary uses the hijacking samples themselves to poison the target model's training dataset.
We show the result of using CelebA as the hijacking dataset and CIFAR-10 as the original dataset in \autoref{fig:CelebAtoCIFAR_tSNE}, and vice verse in \autoref{fig:CIFARToCelebA_tSNE}.
As both figures show, the hijacked samples (\secondAttack) are more clustered with the original samples (Original) than the non-camouflaged hijacking samples (Naive).
Comparing the t-SNE results from the {\firstAttack} attack (\autoref{fig:tsne1stAttk}) and the {\secondAttack} attack (\autoref{fig:tsne2ndAttk}) can further confirms that using the CelebA or CIFAR-10 classification tasks as the hijacking ones are indeed harder than using MNIST; as the original samples in \autoref{fig:tsne1stAttk} are more distant from the rest, compared to the ones in \autoref{fig:tsne2ndAttk}.

Finally, we visualize randomly sampled camouflaged samples for both cases of using the CelebA/CIFAR-10 dataset to hijack a CIFAR-10/CelebA classification task in~\autoref{fig:2ndAtkCamoCelebA}/\autoref{fig:2ndAtkCamoCIFAR}.
As the figures show, our camouflaged samples look visually similar to the original samples with some added artifacts from the hijacking.

\mypara{Discussion of Both Attacks}
As demonstrated in this and the previous section (\autoref{sec:1stAttkeval}), both of our model hijacking attacks, i.e., the {\firstAttack} and {\secondAttack} attacks, achieve strong performance, i.e., they achieve a comparable Attack Success Rate when compared to the Naive attack, and similar utility compared to the clean models.
Moreover, when the hijacking and original datasets are distinct, it is enough to use the {\firstAttack} attack.
However, when both datasets are more complex and similar, then the {\secondAttack} attack is needed to enhance the performance the model hijacking attack.

%-------------------------------------------------------------------------------
\subsection{Hyperparameters}
%-------------------------------------------------------------------------------

We now explore some of the hyperparameters of our model hijacking attacks.
We start by exploring the generalizability of our attack when using different target models and feature extractors.
Next, we evaluate using different loss functions and the transferability of the {\camM}.
Finally, we explore the effects of varying the hijackee dataset size and the poisoning rate on the model hijacking attack.

%-------------------------------------------------------------------------------
\subsubsection{Different Target Models}
\label{sec:diffTargetModel}
%-------------------------------------------------------------------------------

We first evaluate the generalizability of our model hijacking attack on different target models.
We use the CIFAR-10 dataset as our original dataset and evaluate the {\firstAttack} and  {\secondAttack} attacks using MNIST and CelebA as the hijacking datasets, respectively.

We follow the previously mentioned setup in \autoref{sec:1stAttkeval} and \autoref{sec:2ndAttkeval} to implement the {\firstAttack} and {\secondAttack} attack, with the exception of using GoogLeNet~\cite{SLJSRAEVR15} and VGG16~\cite{SZ15} as the target models.
To accelerate convergence, we use pretrained versions of the target models.

\begin{figure}[!t]
\centering
\begin{subfigure}{0.49\columnwidth}
\includegraphics[width=\columnwidth]{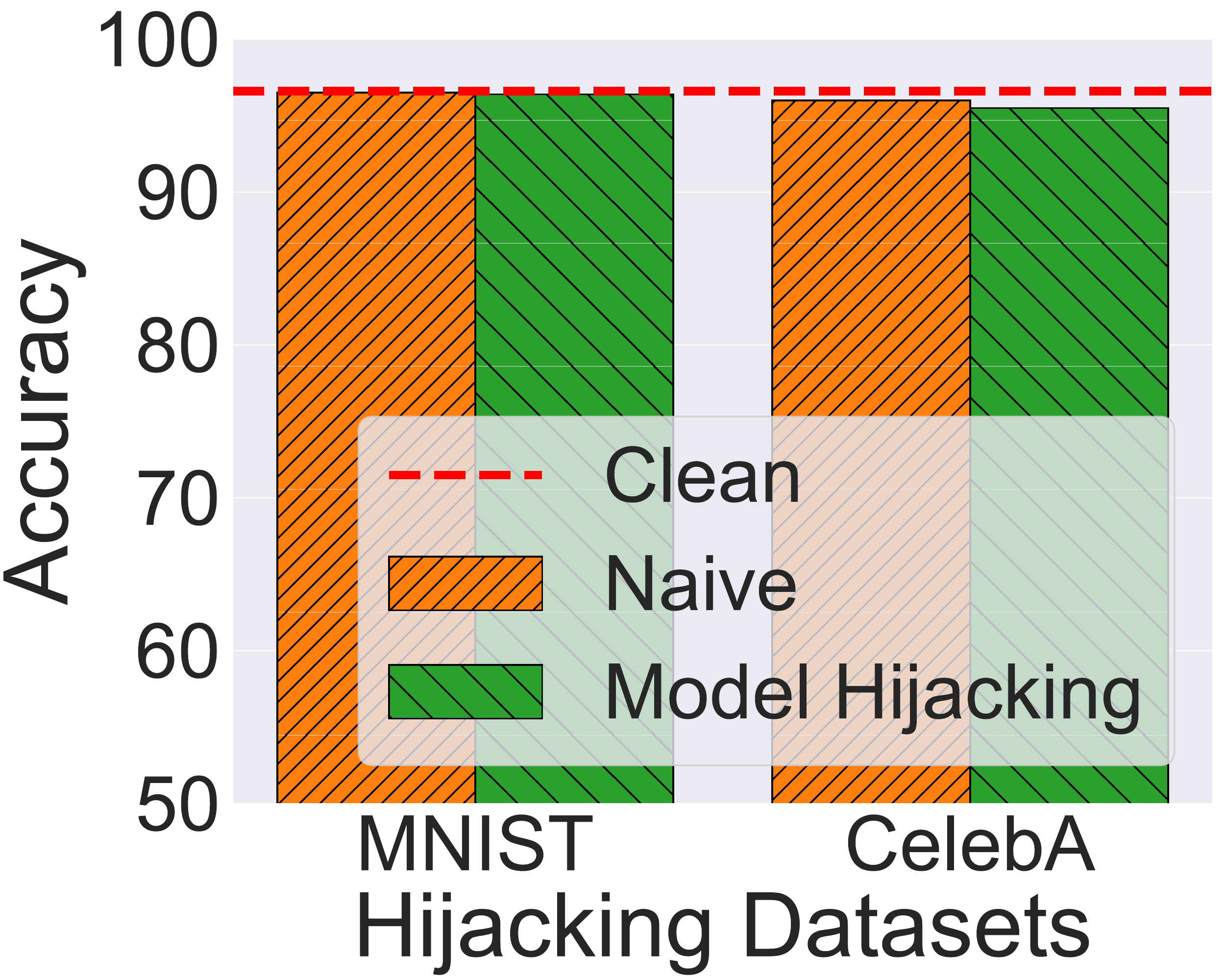}
\caption{Utility (GoogLeNet)}
\label{fig:utlityGoogLeNet} 
\end{subfigure}
\begin{subfigure}{0.49\columnwidth}
\includegraphics[width=\columnwidth]{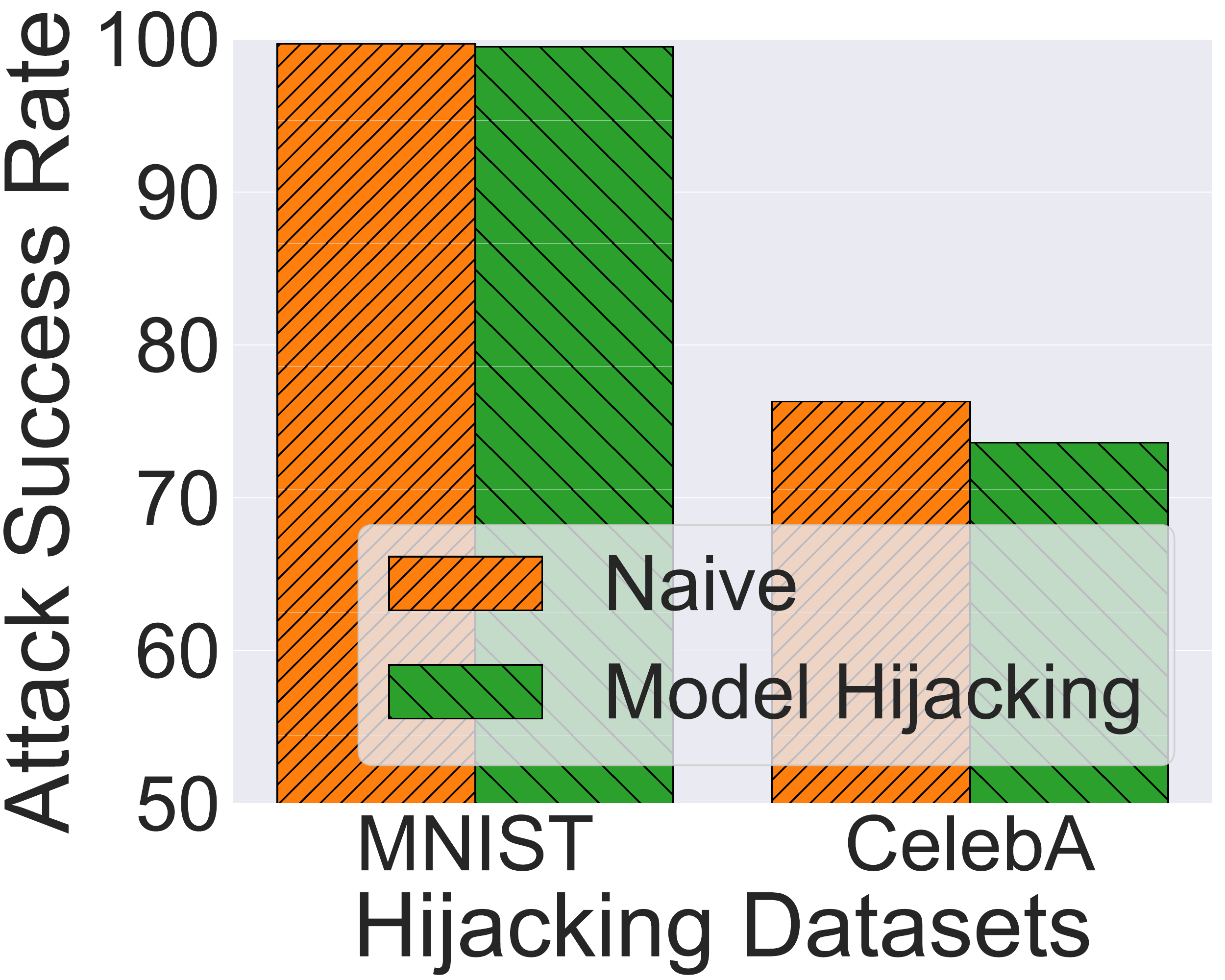}
\caption{ASR (GoogLeNet)}
\label{fig:ASRGoogLeNet} 
\end{subfigure}
\begin{subfigure}{0.49\columnwidth}
\includegraphics[width=\columnwidth]{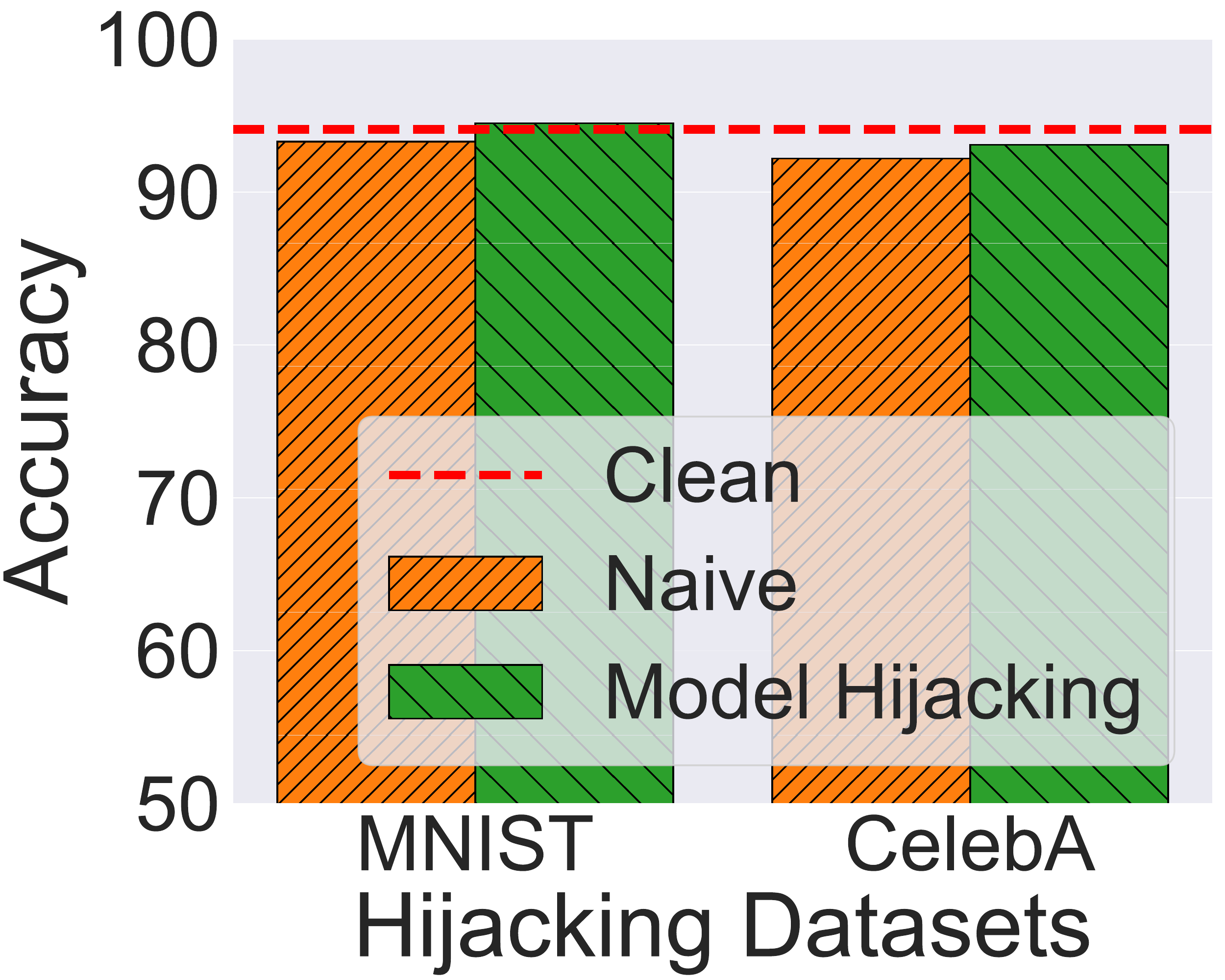}
\caption{Utility (VGG)}
\label{fig:utlityVGG} 
\end{subfigure}
\begin{subfigure}{0.49\columnwidth}
\includegraphics[width=\columnwidth]{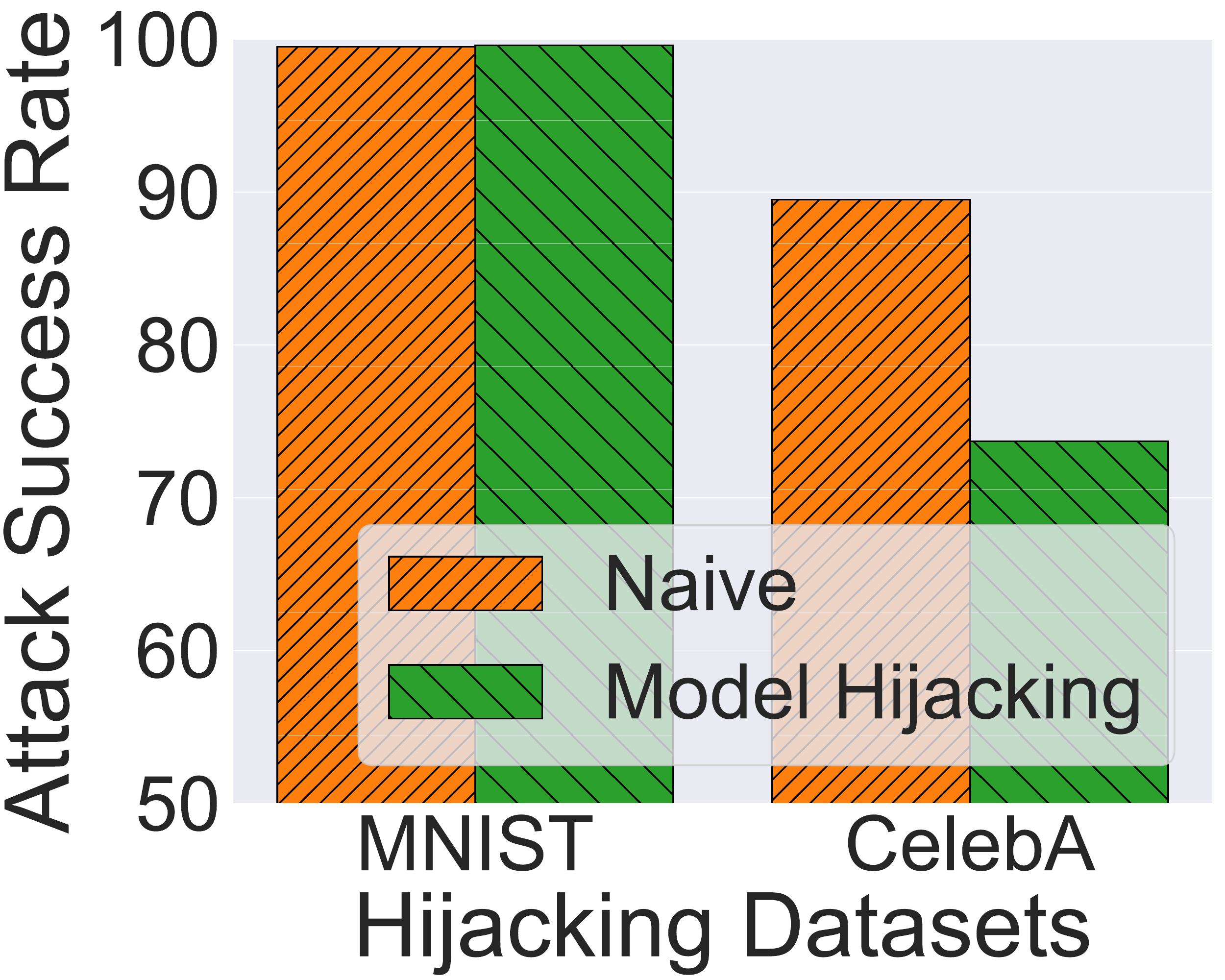}
\caption{ASR (VGG)}
\label{fig:ASRVGG} 
\end{subfigure}
\caption{The results of the {\firstAttack} and {\secondAttack} attacks when targeting a GoogLeNet (\autoref{fig:utlityGoogLeNet} and \autoref{fig:ASRGoogLeNet}) and VGG (\autoref{fig:utlityVGG} and \autoref{fig:ASRVGG}) based models. 
The hijacking datasets are denoted on the x-axis (MNIST for the {\firstAttack} attack and CelebA for the  {\secondAttack} attack) and
the original dataset is CIFAR-10.
Moreover, we show the performance of the clean model using a red dashed line to compare the utility of both attacks.}
\label{fig:diffTargetModel}
\end{figure}

\autoref{fig:diffTargetModel} shows the results for both target models.
As the figure shows, both of our model hijacking attacks, i.e., {\firstAttack} and {\secondAttack}, achieves strong performance against the GoogLeNet and VGG target models.
For instance, both attacks achieves a very similar accuracy similar to a clean model, i.e., the difference is less than 0.5\% and 1\% for the {\firstAttack} and {\secondAttack}, respectively; while achieving high ASR, i.e., above 99\% for MNIST and 70\% for CelebA.

Finally, compared to the Naive attack, our attack achieves similar performance, except for the CelebA case on the VGG target model.
However, it is important to note that for this case, our attack achieves an improved performance with respect to utility.

These results show the generalizability of our model hijacking attack across different target models.

%-------------------------------------------------------------------------------
\subsubsection{Different Feature Extractor}
%-------------------------------------------------------------------------------

Second, we explore using different feature extractor to build the {\camM} and execute the model hijacking attack.
To this end, we use the same evaluation setup similar to \autoref{sec:diffTargetModel} with the exception of using the same target models as \autoref{sec:1stAttkeval} and using the MnasNet~\cite{TCPVSHL19} as the feature  extractor.
We select the MnasNet as it is significantly faster than the MobileNetV2 model.
Finally, we present our results for using a pretrained MnasNet as our feature extractor in ~\autoref{table:featExtract}.

As \autoref{table:featExtract} shows, using the MnasNet achieves good performance for both attacks ({\firstAttack} and {\secondAttack}).
For instance, using the {\firstAttack} attack, the hijacked model achieves 95.2\% accuracy on the original dataset (Utility) and 80.7\% accuracy on the hijacking one (ASR).

This shows the ability of the model hijacking attack to use different models as the feature extractor.
As expected, using different models as the feature extractor can have different effects on the final performance of the model hijacking attack.
However, we believe the model hijacking attack will have a strong attack performance as far as the feature extractor used has acceptable performance.
We plan -- in future work -- to try using multiple feature extractors while training the {\camM} to further increase the independence of the model hijacking attack from the underlying feature extractor used.

\begin{table}[h!]
\centering
\caption{The performance of the {\firstAttack} and {\secondAttack} attacks using MNIST and CelebA as hijacking datasets to attack a CIFAR-10 classification model, while using MnasNet as the Feature  Extractor.}
\label{table:featExtract}
\begin{tabular}{lcc} 
\toprule
Hijacking Dataset & Utility  & Attack Success Rate  \\ 
\midrule
MNIST & 95.2 & 80.7 \\
CelebA & 92.8 & 60.5\\
\bottomrule
\end{tabular}
\end{table}

%-------------------------------------------------------------------------------
\subsubsection{Different Loss Functions}
%-------------------------------------------------------------------------------

Next, we evaluate using different loss functions to implement our model hijacking attack.
More concretely, we evaluate the effect of using the L2 instead of the L1 distance to implement our attack.
We follow the same evaluation setup as \autoref{sec:evalSettings} with the exception
of using L2 instead of L1 distance.

Our experiments show that using L2 distance achieves a strong performance as presented in ~\autoref{table:L2Loss}.
For instance, using the CelebA dataset as the hijacking one to attack a CIFAR-10 classification model results in 86.1\% accuracy and 63.2\% ASR.
This constitutes a drop in performance compared to when using the L1 loss with approximately 10\% for the ASR, however, it improves the accuracy by $6.1\%$.

\begin{table}[h!]
\centering
\caption{The performance of the {\firstAttack} and {\secondAttack} attacks using MNIST and CelebA as hijacking datasets to attack a CIFAR-10 classification model, while using L2 instead of L1 distance as the loss function.}
\label{table:L2Loss}
\begin{tabular}{lcc} 
\toprule
Hijacking Dataset & Utility  & Attack Success Rate  \\ 
\midrule
MNIST & 90.1 & 99.5 \\
CelebA & 86.1& 63.2\\
\bottomrule
\end{tabular}
\end{table}

%-------------------------------------------------------------------------------
\subsubsection{Transferability of the {\camM}}
\label{sec:CamTransferability}
%-------------------------------------------------------------------------------

We now evaluate the transferability of the {\camM}.
To this end, we use the previously trained {\camM}s used in \autoref{sec:1stAttkeval} and \autoref{sec:2ndAttkeval} to hijack a CIFAR-100 classification model with MNIST and CelebA as the hijacking datasets, respectively.
We use the pretrained {\camM}s to implement the {\firstAttack} and {\secondAttack} attacks as previously introduced in ~\autoref{sec:AttackMeth} and \autoref{sec:advAttackMeth}, respectively.

Our experiments show that the {\firstAttack} attack achieves 81.8\% accuracy with a 99.5\% ASR for the MNIST hijacking dataset.
Similarly, the {\secondAttack} attack achieves 78.6\% accuracy with a 76.3\% ASR for the CelebA hijacking dataset.

These results further demonstrate the transferability of the {\camM} after its training.
In other words, the adversary can train a {\camM} and use it to hijack different models with different classification tasks.

%-------------------------------------------------------------------------------
\subsubsection{Hijackee Dataset Size}
%-------------------------------------------------------------------------------

We now evaluate the effect of the size of the hijackee dataset.
Here, we use our {\secondAttack} attack to hijack a CIFAR-10 classification model with the CelebA dataset.

We evaluate a range of different sizes for the hijackee dataset, namely we set the size to $10,100,1,000,$ and $10,000$ samples.
For each setting, we hijack the CIFAR-10 model and calculate both metrics,i.e., Utility and Attack Success rate.

Executing the {\secondAttack} attack using hijackee dataset with the size of $10$, $100$, $1,000$ and $10,000$ achieves $44.7\%$, $60.5\%$, $73.7$ and $65.8\%$ Attack Success Rate, with an accuracy of $82.4\%$, $87.4\%$, $85.9\%$ and $87.4\%$, respectively.
As the results show, using a hijackee dataset of 10 samples is too small for executing the model hijacking attack.
However, setting the size to 100 samples or more is already enough for the {\secondAttack} hijacking attack.
We select the hijackee with a size $1,000$ as it achieves the best overall performance compared to the other two.

%-------------------------------------------------------------------------------
\subsubsection{Poisoning Rate}
%-------------------------------------------------------------------------------

Next, we evaluate the effect of varying the poisoning rate on our model hijacking attack.
In other words, we use different sizes of the hijacking dataset.
To this end, we evaluate both of our {\firstAttack} and {\secondAttack} attacks while setting the size of the hijacking dataset (poisoning data) from $10,000$ to $40,000$ with a step of $10,000$.
For both attacks, we set the original task to CIFAR-10 classification.
For the hijacking task, we use MNIST classification for the {\firstAttack} attack and CelebA classification for the {\secondAttack} attack.

Our results show that for the simple case of using MNIST as the hijacking dataset, $10,000$, i.e., a poisoning rate of 17\%, hijacking samples are enough for our {\firstAttack} attack to hijack a CIFAR-10 classification model.
More concretely, hijacking the model with $10,000$ hijacking samples results in the same Attack Success Rate ($99\%$) as the hijacked model with $40,000$ samples, similarly, the difference between the utility of both models is negligible.

However, for the more complex task of using CelebA as the hijacking dataset to execute the {\secondAttack} attack and hijack a CIFAR-10 classification model; the size of the hijacking dataset has a significant effect.
For instance, the Attack Success rate is reduced from $73.7\%$ to only $63.2\%$ when using $20,000$ samples, i.e., a poisoning rate of 28\%, instead of $40,000$.
However, since there are fewer hijacking samples, the Utility of the hijacked model increases from $85.9\%$ to $86.7\%$.

%-------------------------------------------------------------------------------
\section{Related Works}
\label{sec:related}
%-------------------------------------------------------------------------------

In this section, we review some of the related works.
We divide the related works into training and testing time attacks against machine learning models.
We start with the testing times attacks, then the training time ones.

%-------------------------------------------------------------------------------
\subsection{Testing Time Attacks}
%-------------------------------------------------------------------------------

Testing time attacks are the attacks executed by the adversary after the training of the model. 
We briefly review some of the related test time attacks against machine learning models.

\mypara{Adversarial Reprogramming}
One similar attack to our model hijacking attack is adversarial reprogramming~\cite{EGS19}.
Adversarial reprogramming is a test time attack, where the adversary optimizes a program to let the target model perform a different task.
This program itself is an image with the target image padded inside to create the final input to the target model.
The specially crafted input is then inputted to the target model, which performs the different task of classifying the padded image.
The major difference between the adversarial reprogramming attack and our model hijacking attack is the different assumptions of the attacks, i.e., our model hijacking attack does not make any assumption about the target model and is a training time attack, but the adversarial reprogramming is a test time attack in which the adversary assumes knowledge of the target model similar to the assumptions needed for adversarial examples.
Moreover, in the adversarial reprogramming attack, if the program is known, then the model can be easily patched since all images use the same program.
However, for our model hijacking attack, the knowledge of any hijacked sample does not transfer to any other hijacked samples, i.e., there is no common feature/program for the hijacked samples.

\mypara{Adversarial Examples}
Adversarial examples~\cite{PMGJCS17,VL14,CW17,LV15,TKPGBM17,PMJFCS16,XEQ18} are a testing time attack where the adversary optimizes a noise such that when added to an image it gets misclassified.
There exist two variants of the adversarial examples attack.
The first is targeted adversarial examples, where the noise is optimized to classify the input sample to a specific label.
The second is non-targeted ones, where the noise is optimized just to misclassify the input sample.
Another field of work~\cite{OFS17,JG18,ZHRLPB18,JSBZG19} focus on using adversarial examples to enhancing the users/models privacy.

\mypara{Other Attacks Against Machine Learning Models}
There exist multiple other test time attacks such as: Membership inference~\cite{SSSS17,YGFJ18,SZHBFB19} where the adversary aims at finding if a given sample was used into training the target model or not, Dataset reconstruction~\cite{SBBFZ20} where the adversary tries to reconstruct the updating dataset, Model stealing~\cite{TZJRR16,WG18,OSF19,OASF18} where the adversary tries to steal the target model, i.e., build a model with the same performance as the target model, and Model inversion~\cite{CLEKS19,ZJPWLS20,FJR15,FLJLPR14} where the adversary tries to reconstruct or complete some of the training samples of the target model.

%-------------------------------------------------------------------------------
\subsection{Training Time Attacks}
%-------------------------------------------------------------------------------

Training time attacks are the ones where the adversary executes their attack during or before the training of the target model.
We briefly introduce some of the related training time attacks.

\mypara{Data Poisoning Attack}
Data poisoning attack~\cite{JOBLNL18,SMKID18} is a training time attack where the adversary poisons the training dataset of the target model to compromise the model's utility. 
This poisoning of the training dataset is mostly done by flipping the ground truth of a subset of the dataset, such that the training of the target model fails.
There are multiple works for poisoning different machine learning models/settings such as: Federated Learning~\cite{TTGL20}, Support Vector Machines (SVM)~\cite{BNL12}, Regression Learning~\cite{JOBLNL18}, Node Embeddings~\cite{BG192,STLLXCS18}, Next-Item Recommendation~\cite{ZLDG20}, and Neural Code Completion~\cite{SSTS20}.

It is important to mention that our model hijacking attack can be adapted to any setting vulnerable to the data poisoning attack.

\mypara{Backdoor Attack}
The backdoor attack is another type of training time attacks, where the adversary manipulates the target model's training to backdoor it.
The backdooring behavior is usually assigned with a trigger, which is when inserted in any input sample the target model predicts a specified label.
Gu et al.~\cite{GDG17} introduced BadNets the first backdoor attack against machine learning.
BadNets uses a white square at the coroner of the images as a trigger to misclassify the backdoored inputs to a specific label.
Salem et al.~\cite{SWBMZ20} later proposed dynamic backdoor, where instead of using a fixed trigger, they use a dynamic one.
Another similar attack is the Trojan attack~\cite{LMALZWZ18}.
This attack simplifies the assumptions of the backdoor attack by not assuming the knowledge of any sample from the distribution of the target model's training dataset.
There also exist multiple backdoor attacks attacking against Natural Language Processing (NLP) models~\cite{CSBMSWZ21}, federated learning~\cite{WSRVASLP20}, video recognition \cite{ZMZBCJ20}, transfer Learning \cite{YLZZ19},
and others~\cite{SSP20,LMBL20,RHF20,TDLYH20}

The backdoor attack can be considered a specific instance of the model hijacking attack by considering the classification of the backdoored samples as the hijacking dataset.
However, our model hijacking attack is more general, i.e., it poisons the model to implement a completely different task.

%-------------------------------------------------------------------------------
\section{Discussion}
\label{sec:discuess}
%-------------------------------------------------------------------------------

In this section, we first discuss the limitations of our model hijacking attacks.
Then, we review some of the possible defenses against them.

\mypara{Limitation}
The first limitation of our model hijacking attack is that the hijacking dataset cannot have more number of classes than the original one. 
To address this limitation, we propose to use a more complex hierarchical model hijacking attack with multiple virtual layers of classification tasks.

Intuitively, the adversary would start by grouping the hijacking dataset's classes into $x$ clusters, where $x$ is less than the number of classes of the original dataset.
This constitutes the first layer of the hierarchical attack.
Next, the adversary crafts a different backdoor-like trigger, i.e., a colored square at the corner of the input, for each cluster.
To hijack a model, the adversary would need to poison its dataset with the following:

\begin{itemize}
\item Clean hijacking samples: Camouflaged samples without the triggers, with their corresponding cluster label, i.e., the first layer of the hierarchical attack.
\item Triggered hijacking samples: Camouflaged samples with an added trigger on them. 
This trigger is specific to which cluster this sample is from. The labels of these samples are set to their original ones modulo the target model's number of labels.
\end{itemize}

To execute the attack, the adversary uses the {\camM} to camouflage the sample, before querying it to the model.
Then, depending on the output class, they add the corresponding trigger and query it again to the target model.
We evaluate this hierarchical version using the {\firstAttack} attack with MNIST and CIFAR-10 used as the hijacking and original datasets, respectively.
Our experiments show that the utility of the hijacked model is not significantly affected.
However, the ASR is significantly degraded to be lower than 30\%.
This shows the trade-off between having more labels than the original classification task and the attack performance.
We plan to further explore different techniques -- in future work -- which would overcome this limitation with a better attack performance.

The second limitation of our attack is the visual -- unnatural -- artifacts on the camouflaged images. 
To address this limitation we propose different approaches. 
The first approach is to use a more powerful state-of-the-art autoencoder with more layers. 
However, that will come with the expense of increasing the cost of training the {\camM}. 
A different cheaper approach can be to combine multiple norms, e.g., the $L^2$ norm, when calculating the different losses. 
Moreover, we propose to use a weighting parameter to give more weight to the Visual Loss, hence making the output images more natural. 
Finally, a third approach is to add a discriminative model which penalizes the unnatural look of images. We plan to explore and evaluate these approaches in future work.

Finally, the third limitation of our attack is the cost of training the {\camM}.  
We recap that the {\camM} is only trained once at the start of the model hijacking attack, then is used during the training and after the deployment of the hijacked model. 
Moreover, once the {\camM} is trained, it can be used to hijack multiple target models performing a similar task as shown in \autoref{sec:CamTransferability}. 
However, since the training of the {\camM} can be computationally heavy.
We propose to use a pretrained autoencoder and fine-tune it, which can reduce the training time. Moreover, we plan to explore -- in future work -- adapting few-shot learning techniques to further reduce the training cost of the {\camM}.

\mypara{Possible Defenses}
We now discuss some of the possible defenses against the model hijacking attack.
A naive defense is adding noise to the images before inputting them to the model.
This defense can degrade the attack performance, however, it will also degrade the performance of the original task.
A more complex defense is using an autoencoder or different denoising techniques on the training and testing images.
To evaluate this defense, we train an autoencoder on clean CIFAR-10 data and use it as a denoising step before querying the inputs to the target models hijacked with both the {\firstAttack} and {\secondAttack} attacks.
Our results show that indeed using this step can reduce the ASR to almost random guessing, i.e., 11.1\% for MNIST ({\firstAttack}) and 18.4\% for CelebA ({\secondAttack}).
However, it also significantly reduces the utility of the models.
More concretely, the accuracy drops by 41.6\% and 37.2\% for the CelebA and MNIST datasets, respectively.
We plan to further explore different defense techniques which can provide a better ``defense utility'' trade-off.

Another possible defense is to filter the outputs of the target model based on their entropy.
In other words, the model owner first determines a threshold, and then calculate the entropy of each queried sample.
If the entropy of this sample is above/below the threshold, then the model owner can accept/reject it.
To evaluate this defense, we plot the distribution of the entropy for both clean and camouflaged samples and report the results in the Appendix (\autoref{fig:perp}).
The distributions of both clean and camouflaged samples overlap, which would result in a high false-positive rate.
Another challenge with this approach is determining an appropriate threshold.
As the model owner needs to have access to both clean and camouflaged samples, which is a strong assumption in practice.

\mypara{Generalization to Other Domains} 
As previously mentioned, we focus our model hijacking attack on computer-vision based machine learning models.
However, we believe our attack can be extended to other domains.
The most important requirement for the model hijacking attack is the ability to build a {\camM}.
Intuitively, this means the ability to build an encoder-decoder model to transform the hijacking inputs to ones with similar features as the hijackee inputs.

%-------------------------------------------------------------------------------
\section{Conclusion}
\label{sec:conc}
%-------------------------------------------------------------------------------

The continuous evolution of machine learning models has fueled the demand of including other parties in the training of the models, to be able to train the more complex emerging state-of-the-art models.
An example of such machine learning paradigms is federated learning.
This inclusion of new parties has opened new opportunities for adversaries to attack machine learning models.
More concretely, an adversary can now participate in the training of a target model and manipulate the training process to implement their attack.
This paradigm of machine learning attacks is referred to as the training time attacks.

In this work, we propose a new training  time attack against computer vision based machine learning models namely, the model hijacking attack.
In this attack, the adversary poisons the training dataset of a target model to hijack it into performing a hijacking task.
This new type of attacks can cause severe security and accountability risks.
Since the adversary can now hijack a benign model to perform an illegal or unethical task.
Moreover, the hijacked model's owner can now be framed for the illegal or unethical task their model is capable of.
Another risk of the model hijacking attack is parasitic computing, where the adversary can hijack a public accessible model to implement their private task, for saving the costs of training and maintaining their own model.

We propose two different model hijacking attacks, namely the {\firstAttack} attack and the {\secondAttack} attack.
The {\firstAttack} attack utilizes the Semantic and Visual Losses to hijack the target model, while the {\secondAttack} attack in addition to these two losses, utilizes the Adverse Semantic Loss.

Our results show that indeed both of our model hijacking attacks (the {\firstAttack} and {\secondAttack} attacks) can efficiently hijack machine learning models.
For instance, the {\firstAttack} attack achieves $99\%$ Attack Success Rate on the hijacking task (MNIST classification) with only a utility drop of $0.5\%$ on the original task (CIFAR-10 classification).
Similarly, the {\secondAttack} attack achieves $73.7\%$ Attack Success Rate when hijacking a CIFAR-10 model with a CelebA classification -- hijacking -- task, with a utility drop of only $3.8\%$.

%-------------------------------------------------------------------------------
\section*{Acknowledgment}
%-------------------------------------------------------------------------------

The research leading to these results has received funding from the European Research Council under the European Union's Seventh Framework Programme (FP7/2007-2013)/ ERC grant agreement no. 610150-imPACT, and from the Helmholtz Association within the project ``Trustworthy Federated Data Analytics'' (TFDA) (funding number ZT-I-OO1 4).

%-------------------------------------------------------------------------------
\bibliographystyle{plain}
\bibliography{normal_generated_py3}

\begin{thebibliography}{10}

\bibitem{BNL12}
Battista Biggio, Blaine Nelson, and Pavel Laskov.
\newblock {Poisoning Attacks against Support Vector Machines}.
\newblock In {\em {International Conference on Machine Learning (ICML)}}.
  icml.cc / Omnipress, 2012.

\bibitem{BG192}
Aleksandar Bojchevski and Stephan G{\"u}nnemann.
\newblock {Adversarial Attacks on Node Embeddings via Graph Poisoning}.
\newblock In {\em {International Conference on Machine Learning (ICML)}}, pages
  695--704. PMLR, 2019.

\bibitem{BEGHIIKKMMOPRR19}
Keith Bonawitz, Hubert Eichner, Wolfgang Grieskamp, Dzmitry Huba, Alex
  Ingerman, Vladimir Ivanov, Chlo{\'{e}} Kiddon, Jakub Konecn{\'{y}}, Stefano
  Mazzocchi, H.~Brendan McMahan, Timon~Van Overveldt, David Petrou, Daniel
  Ramage, and Jason Roselander.
\newblock {Towards Federated Learning at Scale: System Design}.
\newblock {\em {CoRR abs/1902.01046}}, 2019.

\bibitem{CLEKS19}
Nicholas Carlini, Chang Liu, {\'U}lfar Erlingsson, Jernej Kos, and Dawn Song.
\newblock {The Secret Sharer: Evaluating and Testing Unintended Memorization in
  Neural Networks}.
\newblock In {\em {USENIX Security Symposium (USENIX Security)}}, pages
  267--284. USENIX, 2019.

\bibitem{CW17}
Nicholas Carlini and David Wagner.
\newblock {Towards Evaluating the Robustness of Neural Networks}.
\newblock In {\em {IEEE Symposium on Security and Privacy (S\&P)}}, pages
  39--57. IEEE, 2017.

\bibitem{CSBMSWZ21}
Xiaoyi Chen, Ahmed Salem, Michael Backes, Shiqing Ma, Qingni Shen, Zhonghai Wu,
  and Yang Zhang.
\newblock {BadNL: Backdoor Attacks Against NLP Models with Semantic-preserving
  Improvements}.
\newblock In {\em {Annual Computer Security Applications Conference (ACSAC)}}.
  ACSAC, 2021.

\bibitem{EGS19}
Gamaleldin~F. Elsayed, Ian~J. Goodfellow, and Jascha Sohl{-}Dickstein.
\newblock {Adversarial Reprogramming of Neural Networks}.
\newblock In {\em {International Conference on Learning Representations
  (ICLR)}}, 2019.

\bibitem{FJR15}
Matt Fredrikson, Somesh Jha, and Thomas Ristenpart.
\newblock {Model Inversion Attacks that Exploit Confidence Information and
  Basic Countermeasures}.
\newblock In {\em {ACM SIGSAC Conference on Computer and Communications
  Security (CCS)}}, pages 1322--1333. ACM, 2015.

\bibitem{FLJLPR14}
Matt Fredrikson, Eric Lantz, Somesh Jha, Simon Lin, David Page, and Thomas
  Ristenpart.
\newblock {Privacy in Pharmacogenetics: An End-to-End Case Study of
  Personalized Warfarin Dosing}.
\newblock In {\em {USENIX Security Symposium (USENIX Security)}}, pages 17--32.
  USENIX, 2014.

\bibitem{GDG17}
Tianyu Gu, Brendan Dolan-Gavitt, and Siddharth Grag.
\newblock {Badnets: Identifying Vulnerabilities in the Machine Learning Model
  Supply Chain}.
\newblock {\em {CoRR abs/1708.06733}}, 2017.

\bibitem{HZRS16}
Kaiming He, Xiangyu Zhang, Shaoqing Ren, and Jian Sun.
\newblock {Deep Residual Learning for Image Recognition}.
\newblock In {\em {IEEE Conference on Computer Vision and Pattern Recognition
  (CVPR)}}, pages 770--778. IEEE, 2016.

\bibitem{JOBLNL18}
Matthew Jagielski, Alina Oprea, Battista Biggio, Chang Liu, Cristina
  Nita-Rotaru, and Bo~Li.
\newblock {Manipulating Machine Learning: Poisoning Attacks and Countermeasures
  for Regression Learning}.
\newblock In {\em {IEEE Symposium on Security and Privacy (S\&P)}}, pages
  19--35. IEEE, 2018.

\bibitem{JG18}
Jinyuan Jia and Neil~Zhenqiang Gong.
\newblock {AttriGuard: A Practical Defense Against Attribute Inference Attacks
  via Adversarial Machine Learning}.
\newblock In {\em {USENIX Security Symposium (USENIX Security)}}, pages
  513--529. USENIX, 2018.

\bibitem{JSBZG19}
Jinyuan Jia, Ahmed Salem, Michael Backes, Yang Zhang, and Neil~Zhenqiang Gong.
\newblock {MemGuard: Defending against Black-Box Membership Inference Attacks
  via Adversarial Examples}.
\newblock In {\em {ACM SIGSAC Conference on Computer and Communications
  Security (CCS)}}, pages 259--274. ACM, 2019.

\bibitem{LV15}
Bo~Li and Yevgeniy Vorobeychik.
\newblock {Scalable Optimization of Randomized Operational Decisions in
  Adversarial Classification Settings}.
\newblock In {\em {International Conference on Artificial Intelligence and
  Statistics (AISTATS)}}, pages 599--607. JMLR, 2015.

\bibitem{LMALZWZ18}
Yingqi Liu, Shiqing Ma, Yousra Aafer, Wen-Chuan Lee, Juan Zhai, Weihang Wang,
  and Xiangyu Zhang.
\newblock {Trojaning Attack on Neural Networks}.
\newblock In {\em {Network and Distributed System Security Symposium (NDSS)}}.
  Internet Society, 2018.

\bibitem{LMBL20}
Yunfei Liu, Xingjun Ma, James Bailey, and Feng Lu.
\newblock {Reflection Backdoor: A Natural Backdoor Attack on Deep Neural
  Networks}.
\newblock In {\em {European Conference on Computer Vision (ECCV)}}, pages
  182--199. Springer, 2020.

\bibitem{LLWT15}
Ziwei Liu, Ping Luo, Xiaogang Wang, and Xiaoou Tang.
\newblock {Deep Learning Face Attributes in the Wild}.
\newblock In {\em {IEEE International Conference on Computer Vision (ICCV)}},
  pages 3730--3738. IEEE, 2015.

\bibitem{OASF18}
Seong~Joon Oh, Max Augustin, Bernt Schiele, and Mario Fritz.
\newblock {Towards Reverse-Engineering Black-Box Neural Networks}.
\newblock In {\em {International Conference on Learning Representations
  (ICLR)}}, 2018.

\bibitem{OFS17}
Seong~Joon Oh, Mario Fritz, and Bernt Schiele.
\newblock {Adversarial Image Perturbation for Privacy Protection -- A Game
  Theory Perspective}.
\newblock In {\em {IEEE International Conference on Computer Vision (ICCV)}},
  pages 1482--1491. IEEE, 2017.

\bibitem{OSF19}
Tribhuvanesh Orekondy, Bernt Schiele, and Mario Fritz.
\newblock {Knockoff Nets: Stealing Functionality of Black-Box Models}.
\newblock In {\em {IEEE Conference on Computer Vision and Pattern Recognition
  (CVPR)}}, pages 4954--4963. IEEE, 2019.

\bibitem{PMGJCS17}
Nicolas Papernot, Patrick~D. McDaniel, Ian Goodfellow, Somesh Jha, Z.~Berkay
  Celik, and Ananthram Swami.
\newblock {Practical Black-Box Attacks Against Machine Learning}.
\newblock In {\em {ACM Asia Conference on Computer and Communications Security
  (ASIACCS)}}, pages 506--519. ACM, 2017.

\bibitem{PMJFCS16}
Nicolas Papernot, Patrick~D. McDaniel, Somesh Jha, Matt Fredrikson, Z.~Berkay
  Celik, and Ananthram Swami.
\newblock {The Limitations of Deep Learning in Adversarial Settings}.
\newblock In {\em {IEEE European Symposium on Security and Privacy (Euro
  S\&P)}}, pages 372--387. IEEE, 2016.

\bibitem{RHF20}
Adnan~Siraj Rakin, Zhezhi He, and Deliang Fan.
\newblock {TBT: Targeted Neural Network Attack with Bit Trojan}.
\newblock In {\em {IEEE Conference on Computer Vision and Pattern Recognition
  (CVPR)}}, pages 13198--13207. IEEE, 2020.

\bibitem{SSP20}
Aniruddha Saha, Akshayvarun Subramanya, and Hamed Pirsiavash.
\newblock {Hidden Trigger Backdoor Attacks}.
\newblock In {\em {AAAI Conference on Artificial Intelligence (AAAI)}}, pages
  11957--11965. AAAI, 2020.

\bibitem{SBBFZ20}
Ahmed Salem, Apratim Bhattacharya, Michael Backes, Mario Fritz, and Yang Zhang.
\newblock {Updates-Leak: Data Set Inference and Reconstruction Attacks in
  Online Learning}.
\newblock In {\em {USENIX Security Symposium (USENIX Security)}}, pages
  1291--1308. USENIX, 2020.

\bibitem{SWBMZ20}
Ahmed Salem, Rui Wen, Michael Backes, Shiqing Ma, and Yang Zhang.
\newblock {Dynamic Backdoor Attacks Against Machine Learning Models}.
\newblock {\em {CoRR abs/2003.03675}}, 2020.

\bibitem{SZHBFB19}
Ahmed Salem, Yang Zhang, Mathias Humbert, Pascal Berrang, Mario Fritz, and
  Michael Backes.
\newblock {ML-Leaks: Model and Data Independent Membership Inference Attacks
  and Defenses on Machine Learning Models}.
\newblock In {\em {Network and Distributed System Security Symposium (NDSS)}}.
  Internet Society, 2019.

\bibitem{SHZZC18}
Mark Sandler, Andrew~G. Howard, Menglong Zhu, Andrey Zhmoginov, and
  Liang{-}Chieh Chen.
\newblock {MobileNetV2: Inverted Residuals and Linear Bottlenecks}.
\newblock In {\em {IEEE Conference on Computer Vision and Pattern Recognition
  (CVPR)}}, pages 4510--4520. IEEE, 2018.

\bibitem{SSTS20}
Roei Schuster, Congzheng Song, Eran Tromer, and Vitaly Shmatikov.
\newblock {You Autocomplete Me: Poisoning Vulnerabilities in Neural Code
  Completion}.
\newblock {\em {CoRR abs/2007.02220}}, 2020.

\bibitem{SHNSSDG18}
Ali Shafahi, W~Ronny Huang, Mahyar Najibi, Octavian Suciu, Christoph Studer,
  Tudor Dumitras, and Tom Goldstein.
\newblock {Poison Frogs! Targeted Clean-Label Poisoning Attacks on Neural
  Networks}.
\newblock In {\em {Annual Conference on Neural Information Processing Systems
  (NeurIPS)}}, pages 6103--6113. NeurIPS, 2018.

\bibitem{SSSS17}
Reza Shokri, Marco Stronati, Congzheng Song, and Vitaly Shmatikov.
\newblock {Membership Inference Attacks Against Machine Learning Models}.
\newblock In {\em {IEEE Symposium on Security and Privacy (S\&P)}}, pages
  3--18. IEEE, 2017.

\bibitem{SZ15}
Karen Simonyan and Andrew Zisserman.
\newblock {Very Deep Convolutional Networks for Large-Scale Image Recognition}.
\newblock In {\em {International Conference on Learning Representations
  (ICLR)}}, 2015.

\bibitem{SMKID18}
Octavian Suciu, Radu M\u{a}rginean, Yi\u{g}itcan Kaya, Hal~Daum\'e III, and
  Tudor Dumitra\c{s}.
\newblock {When Does Machine Learning FAIL? Generalized Transferability for
  Evasion and Poisoning Attacks}.
\newblock In {\em {USENIX Security Symposium (USENIX Security)}}, pages
  1299--1316. USENIX, 2018.

\bibitem{STLLXCS18}
Mingjie Sun, Jian Tang, Huichen Li, Bo~Li, Chaowei Xiao, Yao Chen, and Dawn
  Song.
\newblock {Data Poisoning Attack against Unsupervised Node Embedding Methods}.
\newblock {\em {CoRR abs/1810.12881}}, 2018.

\bibitem{SLJSRAEVR15}
Christian Szegedy, Wei Liu, Yangqing Jia, Pierre Sermanet, Scott~E. Reed,
  Dragomir Anguelov, Dumitru Erhan, Vincent Vanhoucke, and Andrew Rabinovich.
\newblock {Going Deeper with Convolutions}.
\newblock In {\em {IEEE Conference on Computer Vision and Pattern Recognition
  (CVPR)}}, pages 1--9. IEEE, 2015.

\bibitem{TCPVSHL19}
Mingxing Tan, Bo~Chen, Ruoming Pang, Vijay Vasudevan, Mark Sandler, Andrew
  Howard, and Quoc~V. Le.
\newblock {MnasNet: Platform-Aware Neural Architecture Search for Mobile}.
\newblock In {\em {IEEE Conference on Computer Vision and Pattern Recognition
  (CVPR)}}, pages 2820--2828. IEEE, 2019.

\bibitem{TDLYH20}
Ruixiang Tang, Mengnan Du, Ninghao Liu, Fan Yang, and Xia Hu.
\newblock {An Embarrassingly Simple Approach for Trojan Attack in Deep Neural
  Networks}.
\newblock In {\em {ACM Conference on Knowledge Discovery and Data Mining
  (KDD)}}, pages 218--228. ACM, 2020.

\bibitem{TTGL20}
Vale Tolpegin, Stacey Truex, Mehmet~Emre Gursoy, and Ling Liu.
\newblock {Data Poisoning Attacks Against Federated Learning Systems}.
\newblock In {\em {European Symposium on Research in Computer Security
  (ESORICS)}}, pages 480--501. Springer, 2020.

\bibitem{TKPGBM17}
Florian Tram{\`e}r, Alexey Kurakin, Nicolas Papernot, Ian Goodfellow, Dan
  Boneh, and Patrick McDaniel.
\newblock {Ensemble Adversarial Training: Attacks and Defenses}.
\newblock In {\em {International Conference on Learning Representations
  (ICLR)}}, 2017.

\bibitem{TZJRR16}
Florian Tram{\`e}r, Fan Zhang, Ari Juels, Michael~K. Reiter, and Thomas
  Ristenpart.
\newblock {Stealing Machine Learning Models via Prediction APIs}.
\newblock In {\em {USENIX Security Symposium (USENIX Security)}}, pages
  601--618. USENIX, 2016.

\bibitem{MH08}
Laurens van~der Maaten and Geoffrey Hinton.
\newblock {Visualizing Data using t-SNE}.
\newblock {\em {Journal of Machine Learning Research}}, 2008.

\bibitem{VL14}
Yevgeniy Vorobeychik and Bo~Li.
\newblock {Optimal Randomized Classification in Adversarial Settings}.
\newblock In {\em {International Conference on Autonomous Agents and
  Multi-agent Systems (AAMAS)}}, pages 485--492. IFAAMAS/ACM, 2014.

\bibitem{WG18}
Binghui Wang and Neil~Zhenqiang Gong.
\newblock {Stealing Hyperparameters in Machine Learning}.
\newblock In {\em {IEEE Symposium on Security and Privacy (S\&P)}}, pages
  36--52. IEEE, 2018.

\bibitem{WSRVASLP20}
Hongyi Wang, Kartik Sreenivasan, Shashank Rajput, Harit Vishwakarma, Saurabh
  Agarwal, Jy~yong Sohn, Kangwook Lee, and Dimitris Papailiopoulos.
\newblock {Attack of the Tails: Yes, You Really Can Backdoor Federated
  Learning}.
\newblock In {\em {Annual Conference on Neural Information Processing Systems
  (NeurIPS)}}. NeurIPS, 2020.

\bibitem{XEQ18}
Weilin Xu, David Evans, and Yanjun Qi.
\newblock {Feature Squeezing: Detecting Adversarial Examples in Deep Neural
  Networks}.
\newblock In {\em {Network and Distributed System Security Symposium (NDSS)}}.
  Internet Society, 2018.

\bibitem{YLZZ19}
Yuanshun Yao, Huiying Li, Haitao Zheng, and Ben~Y. Zhao.
\newblock {Latent Backdoor Attacks on Deep Neural Networks}.
\newblock In {\em {ACM SIGSAC Conference on Computer and Communications
  Security (CCS)}}, pages 2041--2055. ACM, 2019.

\bibitem{YGFJ18}
Samuel Yeom, Irene Giacomelli, Matt Fredrikson, and Somesh Jha.
\newblock {Privacy Risk in Machine Learning: Analyzing the Connection to
  Overfitting}.
\newblock In {\em {IEEE Computer Security Foundations Symposium (CSF)}}, pages
  268--282. IEEE, 2018.

\bibitem{ZLDG20}
Hengtong Zhang, Yaliang Li, Bolin Ding, and Jing Gao.
\newblock {Practical Data Poisoning Attack against Next-Item Recommendation}.
\newblock {\em {CoRR abs/2004.03728}}, 2020.

\bibitem{ZHRLPB18}
Yang Zhang, Mathias Humbert, Tahleen Rahman, Cheng-Te Li, Jun Pang, and Michael
  Backes.
\newblock {Tagvisor: A Privacy Advisor for Sharing Hashtags}.
\newblock In {\em {The Web Conference (WWW)}}, pages 287--296. ACM, 2018.

\bibitem{ZJPWLS20}
Yuheng Zhang, Ruoxi Jia, Hengzhi Pei, Wenxiao Wang, Bo~Li, and Dawn Song.
\newblock {The Secret Revealer: Generative Model-Inversion Attacks Against Deep
  Neural Networks}.
\newblock In {\em {IEEE Conference on Computer Vision and Pattern Recognition
  (CVPR)}}, pages 250--258. IEEE, 2020.

\bibitem{ZMZBCJ20}
Shihao Zhao, Xingjun Ma, Xiang Zheng, James Bailey, Jingjing Chen, and Yu-Gang
  Jiang.
\newblock {Clean-Label Backdoor Attacks on Video Recognition Models}.
\newblock In {\em {IEEE Conference on Computer Vision and Pattern Recognition
  (CVPR)}}, pages 14443--144528. IEEE, 2020.

\end{thebibliography}
%-------------------------------------------------------------------------------

\newpage
%-------------------------------------------------------------------------------
\appendix
%-------------------------------------------------------------------------------

\begin{table*}[!t]
\centering
\caption{The description of the different datasets used in the model hijacking attack.}
\label{table:datasetDesc}
\begin{tabular}{p{0.2\linewidth}  p{0.65\linewidth}} 
\toprule
Dataset & Description  \\ 
\midrule
Original/Target Dataset& The dataset intended to be used by the target model's owner to train their model (the target model).\\
Hijackee Dataset&  The dataset used to camouflage the hijacking samples, i.e., transform the visual appearance of the hijacking samples to ideally
look alike the original dataset or make them harder to detect in general.\\
Hijacking Dataset&  The dataset intended to be used by the adversary to hijack the target model.\\
Camouflaged Dataset& The hijacking dataset after being camouflaged by the {\camM}.\\
Poisoned Dataset& The dataset the model will train with, i.e., the concatenation of the camouflaged and the original datasets.\\
\bottomrule
\end{tabular}
\end{table*}

\begin{figure*}[!t]
\centering
\begin{subfigure}{0.49\columnwidth}
\includegraphics[width=\columnwidth]{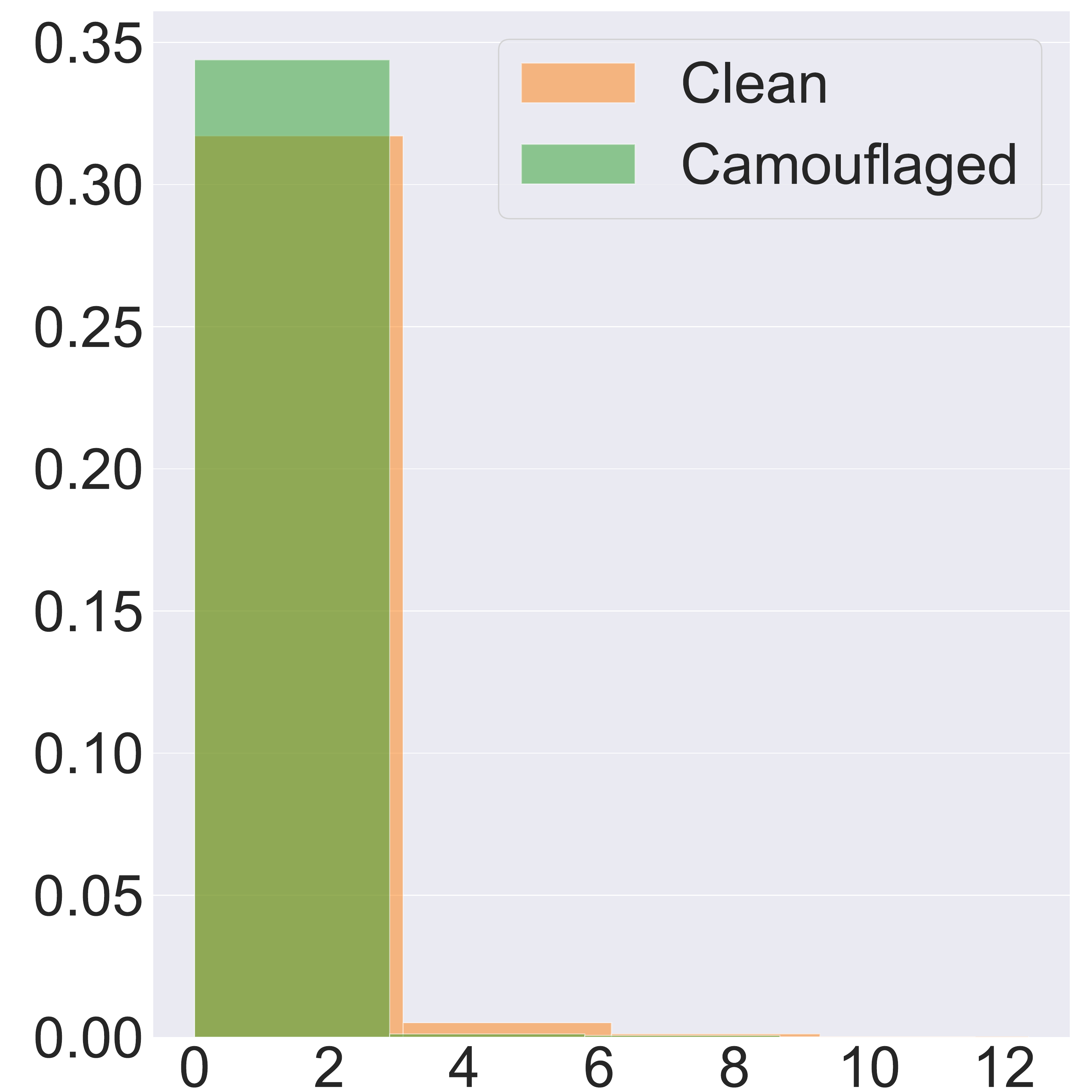}
\caption{MNIST}
\label{fig:perpMNIST} 
\end{subfigure}
\begin{subfigure}{0.49\columnwidth}
\includegraphics[width=\columnwidth]{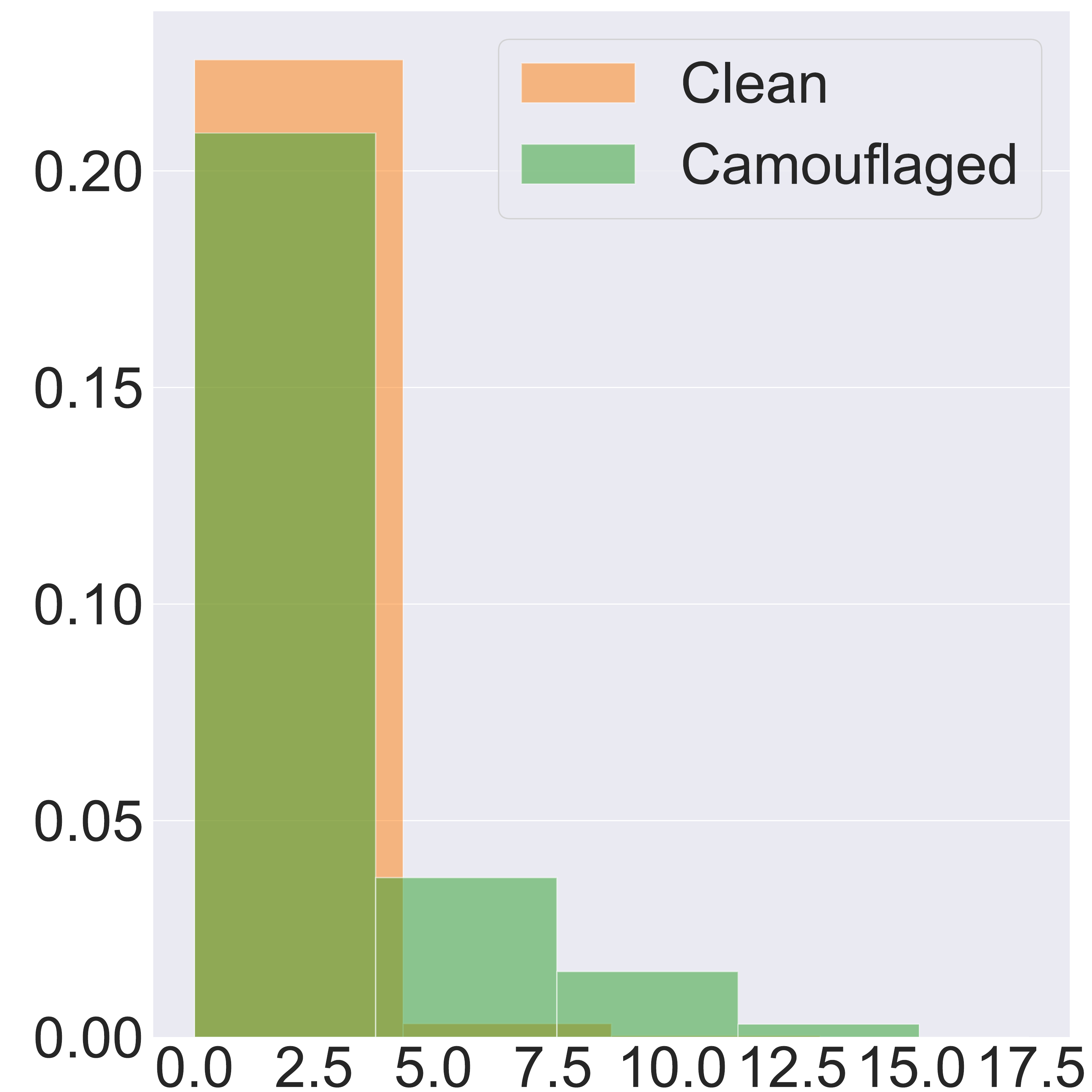}
\caption{CelebA}
\label{fig:perpCelebA} 
\end{subfigure}
\caption{The distribution of the entropy of the outputs of the target model on the clean (CIFAR-10) and camouflaged (MNIST and CelebA) datasets. 
\autoref{fig:perpMNIST} and \autoref{fig:perpCelebA} show the results when hijacking a CIFAR-10 model using MNIST ({\firstAttack} attack) and CelebA ({\secondAttack}), respectively.}
\label{fig:perp}
\end{figure*}

\end{document}